\documentclass[12pt]{article}

\usepackage{graphicx}       

\usepackage[hmargin={2.5cm, 2.5cm}, vmargin={2.5cm, 2.5cm}]{geometry}

\usepackage{hyperref}       
\hypersetup{colorlinks=true, citecolor=black, linkcolor=black, urlcolor=black}

\usepackage{authblk}        

\usepackage{subcaption}      

\usepackage[font=scriptsize, labelfont=bf]{caption} 

\usepackage{sectsty}     
\sectionfont{\normalsize}
\subsectionfont{\small}
\subsubsectionfont{\small}

\usepackage{cite}     

\usepackage{verbatim}     

\usepackage{float}  

\providecommand{\keywords}{\small\textbf{Keywords: }} 

\title{\large \textbf{Detector and physics simulation using heavy ion collisions at NICA-SPD}}

\author[1,a)]{R. Pandey}
\affil[1]{Larsen $\&$ Toubro Limited, Faridabad-121003, Haryana, India.}
\affil[a)]{rishav160999@gmail.com}

\date{}

\begin{document}

\maketitle

\section*{ACKNOWLEDGEMENT}
The author expresses his gratitude to Dr. Igor Denisenko from the Joint Institute for Nuclear Research, Dubna, Russia, for his invaluable support and guidance throughout this research.

\begin{abstract}
The space-time picture of hadron formation in high-energy collisions with nuclear targets is still poorly known. The tests of hadron formation was suggested for the first stage of SPD running. They will require measuring charged pion and proton spectra with the precision better than 10\%. A research has been carried out to check feasibility of such studies at SPD. In this work, $^{12}C-{^{12}C}$ and $^{40}Ca-{^{40}Ca}$ heavy ion collisions at center of mass energy of 11 GeV/nucleon were simulated using the SMASH event generator. Firstly,  the generator-level events were studied. The distribution of track multiplicities and momentum distributions of different types of charged particles were obtained. Secondly, the generated events passed through the full reconstruction using the SpdRoot framework. At this stage, particles were identified using $dE/dx$ measurement and time-of-flight information.  It allowed us to estimate charge track multiplicities in the tracking system and purities of charge particles spectra. The results on multiplicity are important to estimate occupancies in the tracking system, while the results on the pion and proton momentum spectra show that particle identification should be acceptable for validation of hadron formation models. This is the first study of moderate ion collisions for the SPD Experiment.
\end{abstract}

\keywords{Hadron formation effects, Heavy ion collision, SMASH, NICA-SPD, Rapidity, Charged track multiplicity, Particle physics event generator.}

\clearpage

\section{INTRODUCTION}
The SPD detector is primarily optimized to study spin dependent
gluon structure of proton and deuteron using open charm production,
charmonia production, and prompt photons. At the same time, 
its physics program includes studies of various aspects of QCD.
The work is devoted to studies of hadron formation in nuclear
collisions proposed in Ref.~\cite{Abramov:2021vtu}.

\begin{figure}[h]
\centering
\includegraphics[scale=0.35]{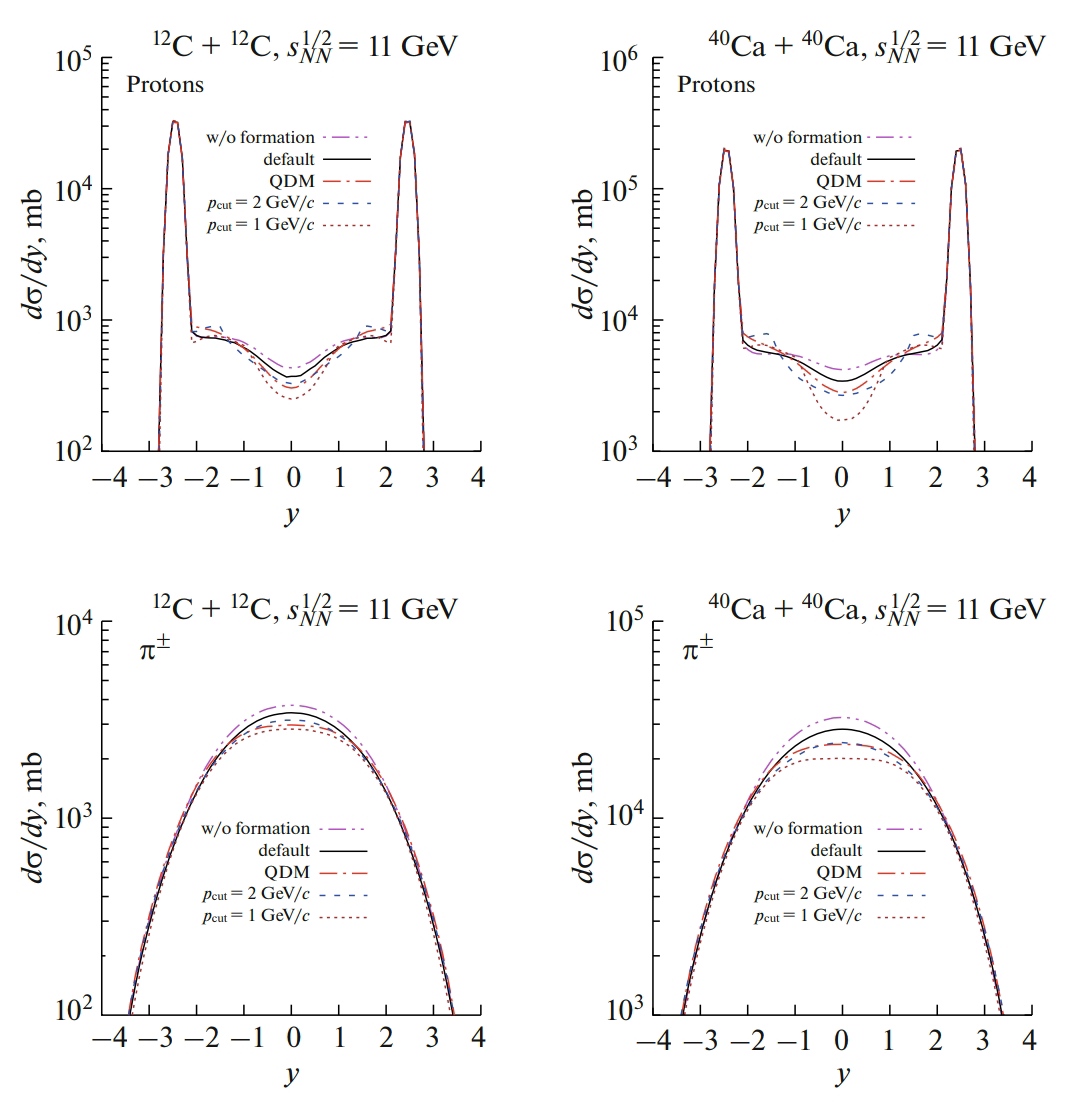}
\caption{Rapidity spectra of protons and charged pions in $^{12}C-{^{12}C}$ and $^{40}Ca-{^{40}Ca}$ collisions.}
\label{ABRAMOV_Paper_Fig.36}
\end{figure}

Hadrons produced in hadron collisions emerge in the form of prehadrons,
which interact with nucleons with reduced strength. This suppression is
poorly known and is described in model dependent way. This suppression
results in different spectra of final particles as is illustrated in
Fig.\ref{ABRAMOV_Paper_Fig.36} for rapidity distributions (in a similar
way it affects the $p_T$ spectrum). Naturally, these spectra
can be used  to study hadron formation effects. The required precision
of such measurements is 10\%.

The aim of this work is to evaluate  feasibility of such measurements
with MC simulation. Here, ion collisions of $^{12}C-{^{12}C}$ and $^{40}Ca-{^{40}Ca}$ at $\sqrt{s}=11 AGeV$ were generated using the SMASH (Simulating Many Accelerated Strongly-interacting Hadrons) event generator. Afterwards, the full simulation and reconstruction was performed using the SpdRoot framework.


\section{NICA FACILITY}
The NICA (Nuclotron based Ion Collider fAcility) collider at Joint Institute for Nuclear Research in Dubna is being built to  provide beams for two experiments. The first experiment, MPD (Multi Purpose Detector), will study properties of dense baryonic matter (matter present at extreme high density in QCD phase diagram) like Quark Gluon Plasma. The second experiment, SPD (Spin Physics Detector), is devoted to study of spin related phonomena and QCD. Once the NICA collider will be operational, scientists will be able to create a special state of matter in laboratory which existed for very short interval of time (\~20$\mu$ sec) just after the big bang. This special state is called as QGP (Quark Gluon Plasma) and it filled the entire universe shortly after the big bang.

\begin{figure}[h]
\centering
\includegraphics[scale=0.3]{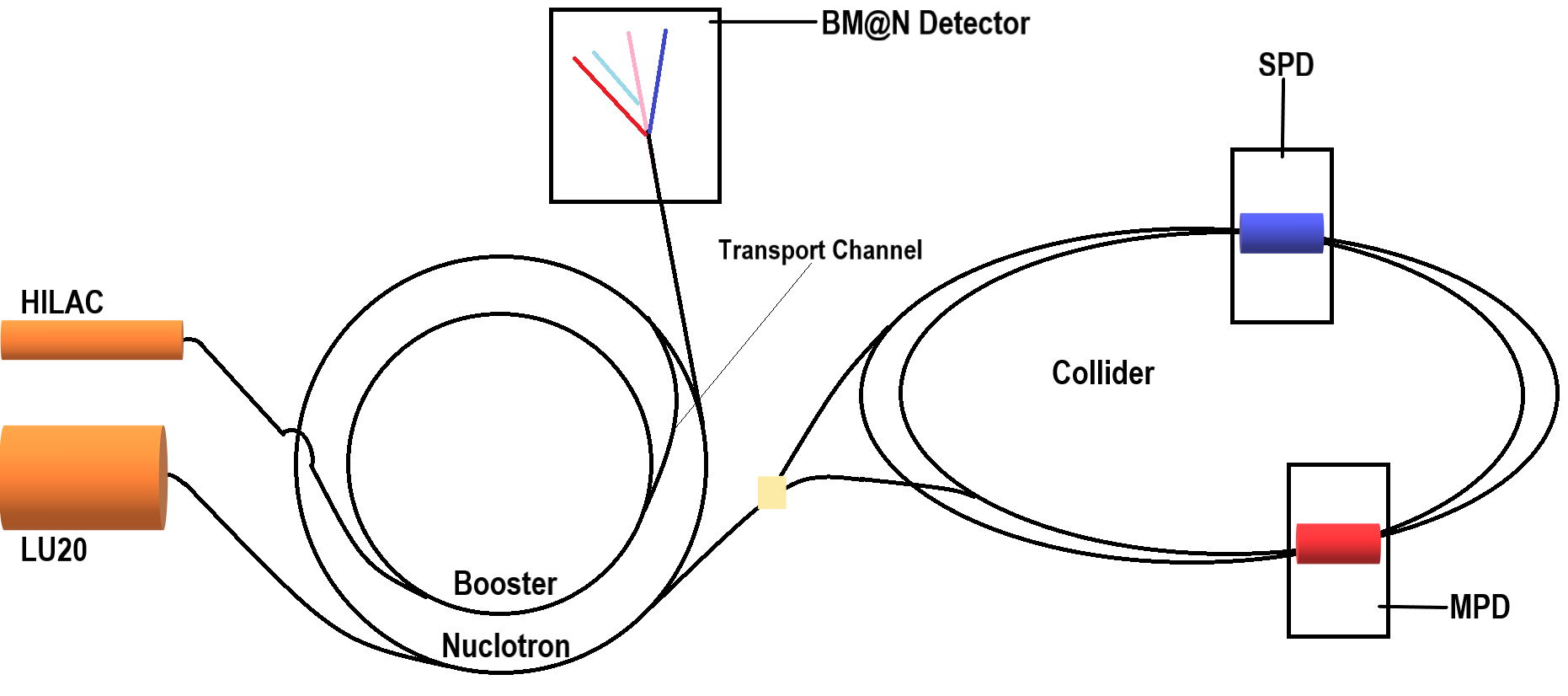}
\caption{Schematic view of NICA complex.}
\label{Schematic view of NICA complex.}
\end{figure}

The main parts of NICA facility consists of two independent injector complex (injector for light ions, and injector for heavy ions-KRION 6T), Light Ion Linear Accelerator (LU20) for accelerating light ions like protons ($H^{+}$), deutrons, and $\alpha$-particles upto 5 MeV of K.E, then Heavy Ion Linear Accelerator (HILAC) to accelerate heavy ions upto Au to a maximum K.E of 3.2 MeV/n, then a Super Conducting (SC) Booster Synchrotron to create ultra high vacuum and to provide complete stripping of heavy ions, then a SC Heavy Ion Synchrotron Nuclotron to accelerate both light and heavy ions to required beam energy. The accelerated beams will collide at two different locations where MPD detector and SPD detector are being built. The schematic view of NICA complex is shown in Fig.\ref{Schematic view of NICA complex.}.

\section{SPD DETECTOR}
\label{SPD DETECTOR}
\begin{figure}[h]
\centering
\includegraphics[width=\textwidth]{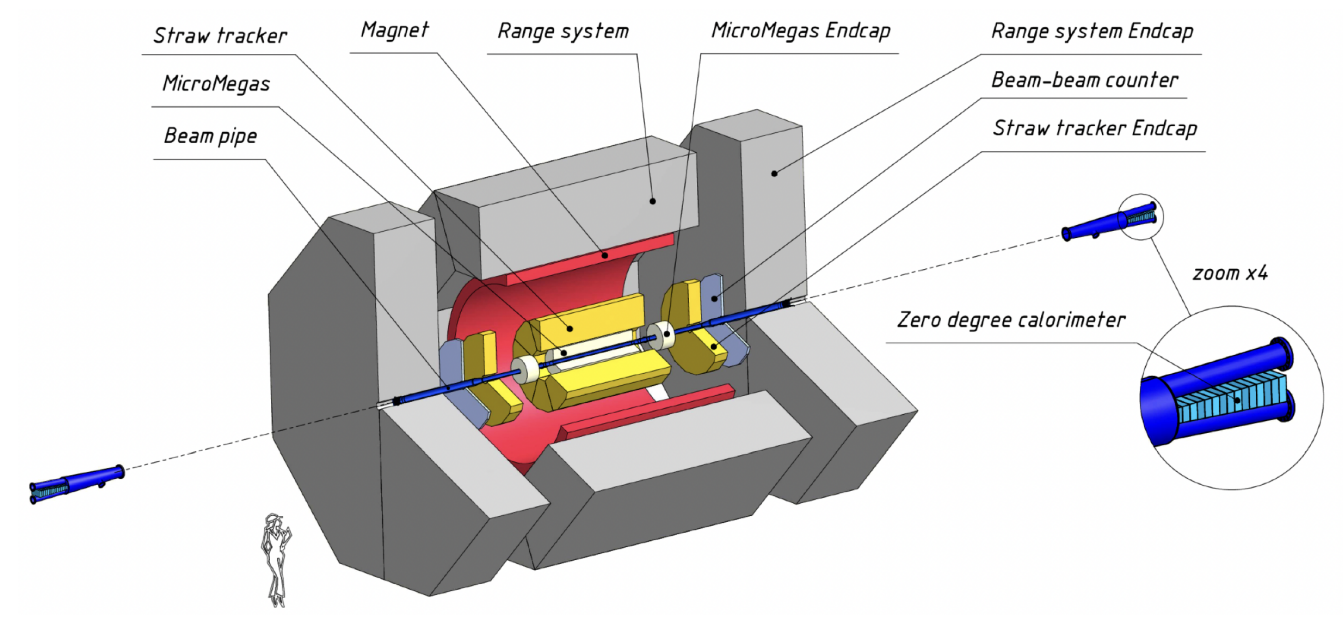}
\caption{Layout of the SPD setup proposed for first stage at NICA-SPD.}
\label{Layout of the SPD setup proposed for first stage at NICA-SPD.}
\end{figure}

The Spin Physics Detector~\cite{SPDproto:2021hnm,TDR} is a $4\pi$ universal detector optimized
to study spin-related phenomena via open charm, charmonia, and promopt photons in the
collisions of polarized p-p or d-d beams with $\sqrt{s_{NN}}$ up to 27 GeV.
However, at first stage of NICA-SPD, the expected collision energy will be from 3.4 up to 10~GeV, and later on after first upgrade, it is expected to reach upto 27 GeV. The general layout depicting isometric projection of SPD setup is shown in Fig.\ref{Layout of the SPD setup proposed for first stage at NICA-SPD.}. The main parts involved in advanced tracking and particle identification capabilities have been shown. (i) The beam pipe passes through the center of the detector, carries the accelerated beams of ions. (ii) The MicroMegas detector is to improves the momentum resolution and tracking efficiency of the tracking system. (iii) The Straw Tracker (ST) detector is for the reconstruction of the primary and secondary particle tracks and for determination of their momenta. (iv) The Time Of Flight (TOF) detector, is a part of Particle Identification (PID) system, and is used for identification of particles like $\pi$, k, and p with long trajectories. (v) The magnet system shown by red color provides $1T$ of magnetic field along the beam axis. This setup is limited to first stage of SPD operation, and will be considered only for the identification of stable charged particles. Neutral particles, like $n^{0}$, photons will be detected at later stages. The main parts of SPD first stage have been explained in detail below. There is a possibility to have TOF system for the first stage studies.

\subsection{CENTRAL TRACKER}
The innermost detector of SPD consists of a MicroMegas-based Central Tracker (MCT). Its purpose is to identify the primary vertex coordinate and to improve momentum resolution and tracking efficiency. It is based on MicroMegas (Micro Mesh Gaseous Structure) technology and detects charged particle by amplifying the charges produced due to ionization of the gas molecules present in detector volume. When an ionizing particle track passes through detector volume, it ionizes the gas molecules and creates few hundreds of $e^{-}$-ion pair. Electrons are accelerated opposite to the direction of applied electric field of $600$~V/cm in ionization gap, while ions are attracted towards cathode. When the $e^{-}$ crosses micromesh, it faces intense electric field ($>30$ KV/cm) and gains enough energy to ionize other gas molecules in its path. During this process an avalanche of $e^{-}$-ion pair is produced ($1e^{-}$ produces $10^4$ $e^{-}$-ion pairs) which is significant to create an electronic signal which is read out by readout electrodes.   

\subsection{STRAW TRACKER}
\label{STRAW TRACKER}
ST is mainly for the reconstruction of primary and secondary particle tracks and measuring their momenta, but also participates in identification of $\pi$, K, and p on via energy deposit $(dE/dx)$ measurements. It consists of two major parts - barrel (covers radius from 270 to 850 mm) and two end-caps. The barrel is divided into 8 modules enclosed in a carbon fiber capsule. Each module has 30 double layers of straw tubes (dia 1cm) which runs parallel (long straw tubes) and perpendicular (short straw tubes) to the beam axis and contains 1500 and 6000 parallel and perpendicular straw tubes respectively. Straw tubes are made of polyethylene terephthalate and outer surface is coated with very thin layer of Cu and Au. Carbon capsule is meant to protect the outer surface of these tubes from humidity. One side and two opposite ends of capsule are provided with small holes where end plugs are fixed. FEE are connected to these end plugs to read the detector signal. Any particle which passes through the long straws will send detector signal to both opposite ends while a particle passing through short straw will send detector signal to any one side of capsule where FEE is attached. Thus, long straws will be read from two opposite ends while short straws will be read from one side. The end-caps of ST are divided into 3 modules and each module has 4 hexadecimal cameras (U, V, X, Y) to record the four coordinates of any physical quantity like four-momentum. The FEE to be used can be similar to the one used at NA64 experiment (for the search of dark matter), or DUNE experiment (to detect and study properties of neutrino).

\subsection{TIME OF FLIGHT DETECTOR}
TOF detector is the part of PID system. Similar to ST, the TOF provides identification of $\pi$, k, and p by measuring their flight time. The energy loss data registered by ST can be used together with the data from TOF for correct identification of particle tracks. The TOF distinguishes charged particles (mainly $\pi$ and k) in the momentum range up to 1.5~GeV. The major parts of TOF comprises of a barrel and two end-caps. For the first stage of NICA-SPD, two different designs of TOF has been suggested. First one is TOF based on multigap timing Resistive Plate Chambers (mRPC), which will consist 220 rectangular plate chambers (160 for the barrel and 30 each for end-caps). Second one is based on Plastic Scintillator Tiles and will comprise 10.1K small scintillator tiles (7.4K for barrel and 1.4K for each end-caps). Scintillator has a property of emitting light in visible region when an ionizing radiation passes through it. So, in this design when a particle passes through TOF, scintillated photons are produced which are detected by four Si Photo Multipliers (SiPMs) present at each sensor board attached at two extreme ends of scintillator tile.

\section{EVENT GENERATION}
$^{12}C-{^{12}C}$ and $^{40}Ca-{^{40}Ca}$ heavy ion collisions at $\sqrt{s}=11$~AGeV with maximum impact parameter set to 8~fm for C-C and 11~fm for Ca-Ca were simulated using SMASH. The fermi motion was assumed to be ``frozen'' and 100K events were generated for each heavy ion collision. The SMASH input file for C-C collision is shown below.
{\small 
\begin{verbatim}
*********** SMASH INPUT ************

config.yaml file for C-C collision.

Logging:
    default: INFO
General:
    Modus:          Collider
    Time_Step_Mode: Fixed
    Delta_Time:     0.1
    End_Time:       200.0
    Randomseed:     -1
    Nevents:        100000
Output:
    Output_Interval: 10.0
    Particles:
        Format:       ["Oscar2013"]
Modi:
    Collider:
        Projectile:
            Particles: {2212: 6, 2112: 6} #C-12
        Target:
            Particles: {2212: 6, 2112: 6} #C-12       
        Sqrtsnn: 11.0        
        Impact:
            Sample: "quadratic"
            Range: [0.0, 8.0]
        Fermi_Motion: "frozen"    
            
************************************
\end{verbatim}
}

\begin{figure}[h]
\centering
\begin{subfigure}[h]{0.49\textwidth}
\centering
\includegraphics[width=\textwidth, height=0.4\textheight]{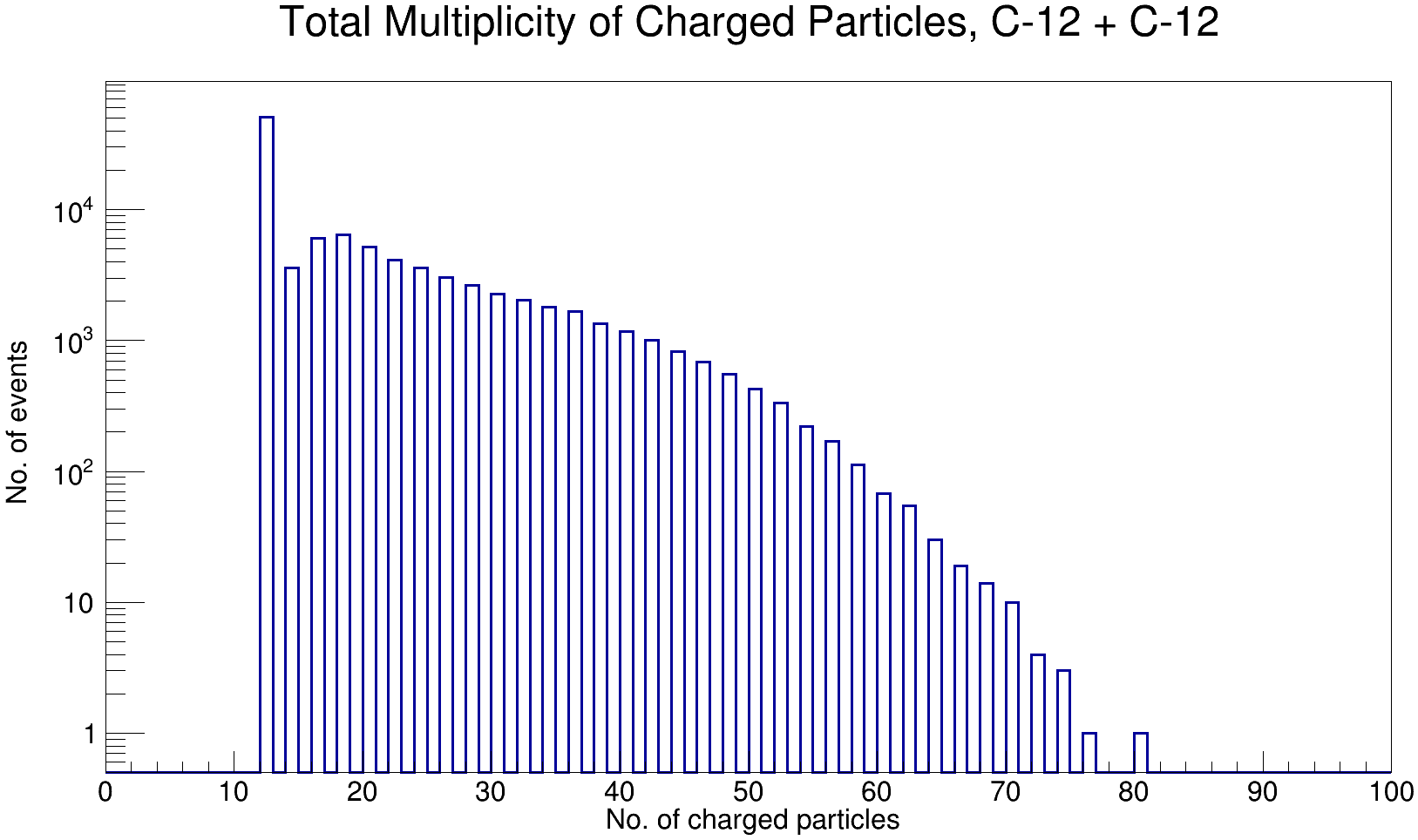}
\caption{}
\label{Total multiplicity of all charged particles C12.}
\end{subfigure}
\hfill
\begin{subfigure}[h]{0.49\textwidth}
	\centering
	\includegraphics[width=\textwidth, height=0.4\textheight]{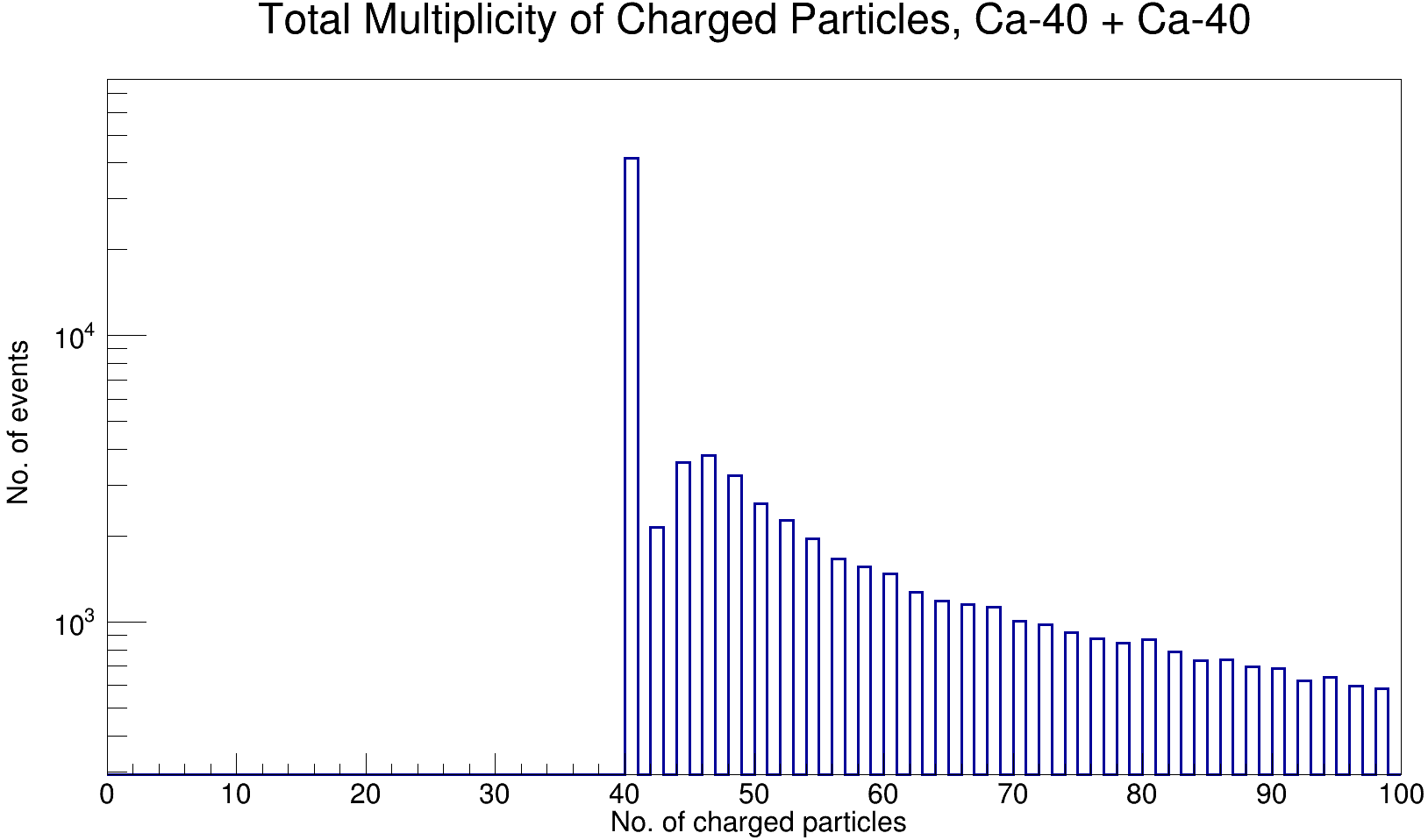}
	\caption{}
	\label{Total multiplicity of all charged particles Ca40.}
\end{subfigure}
\caption{Generator-level multiplicity of charged particles for $^{12}C-{^{12}C}$ collision (a)
	and $^{40}Ca-{^{40}Ca}$ collisions (b).}
\label{gen-multipl}
\end{figure}

\begin{figure}[h]
\centering
\begin{subfigure}[h]{0.49\textwidth}
\centering
\includegraphics[width=\textwidth, height=0.36\textheight]{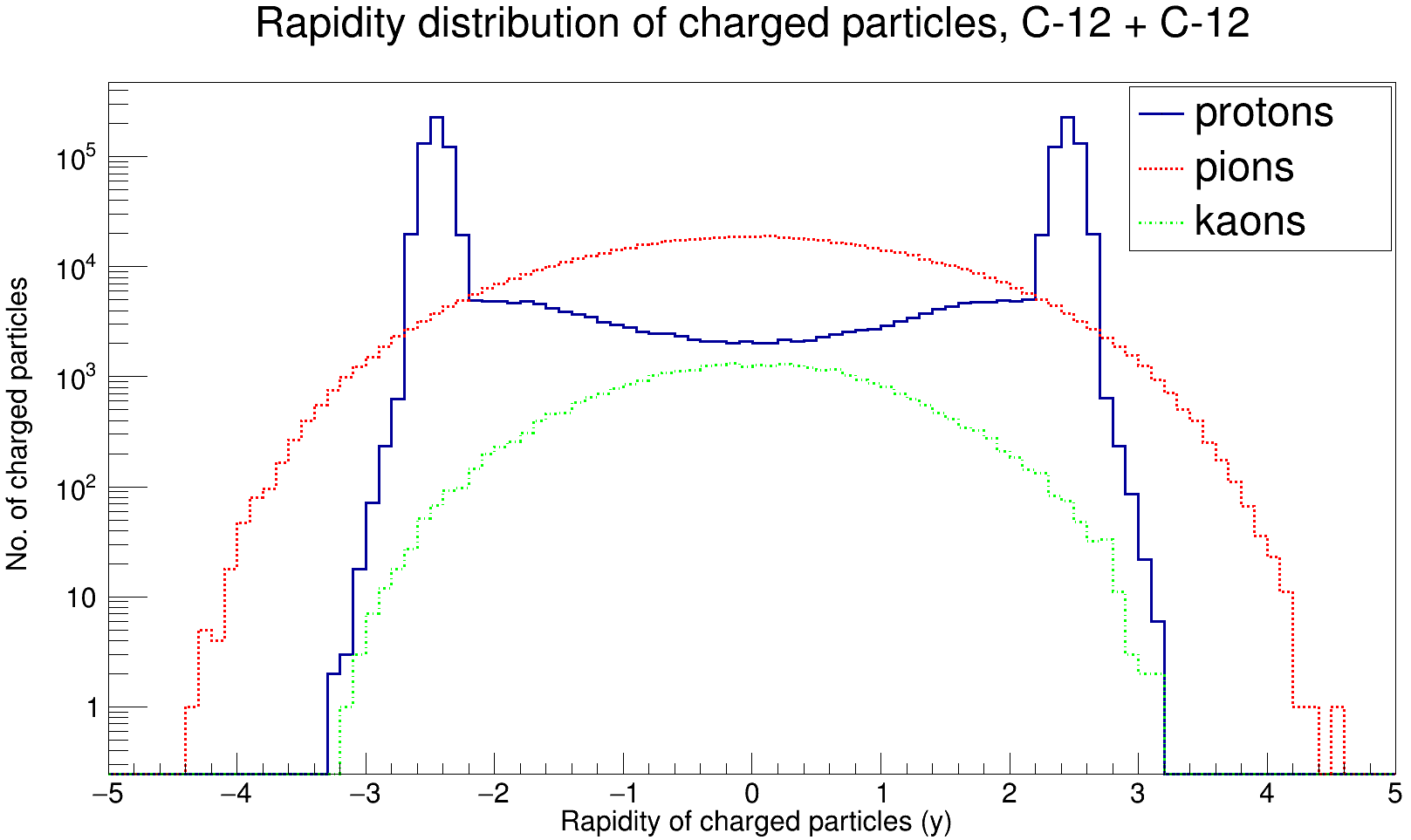}
\caption{}
\end{subfigure}
\hfill
\begin{subfigure}[h]{0.49\textwidth}
\centering
\includegraphics[width=\textwidth, height=0.36\textheight]{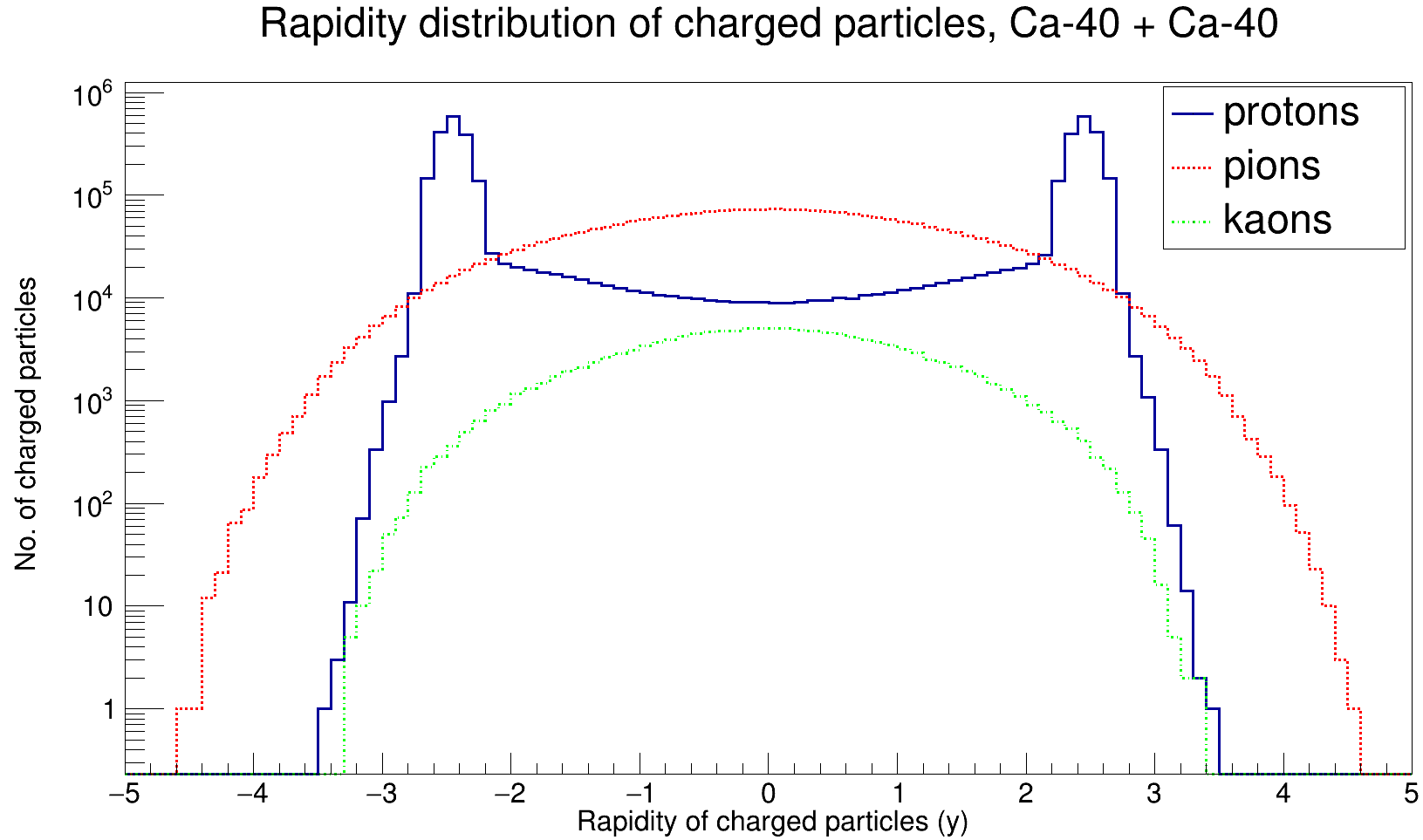}
\caption{}
\end{subfigure}
\caption{Rapidity distribution of charged particles in $^{12}C-{^{12}C}$ (a) and $^{40}Ca-{^{40}Ca}$ (b) collision.}
\label{gen-rapidity}
\end{figure}

\begin{figure}[h]
\centering
\begin{subfigure}[h]{0.49\textwidth}
\centering
\includegraphics[scale=0.14]{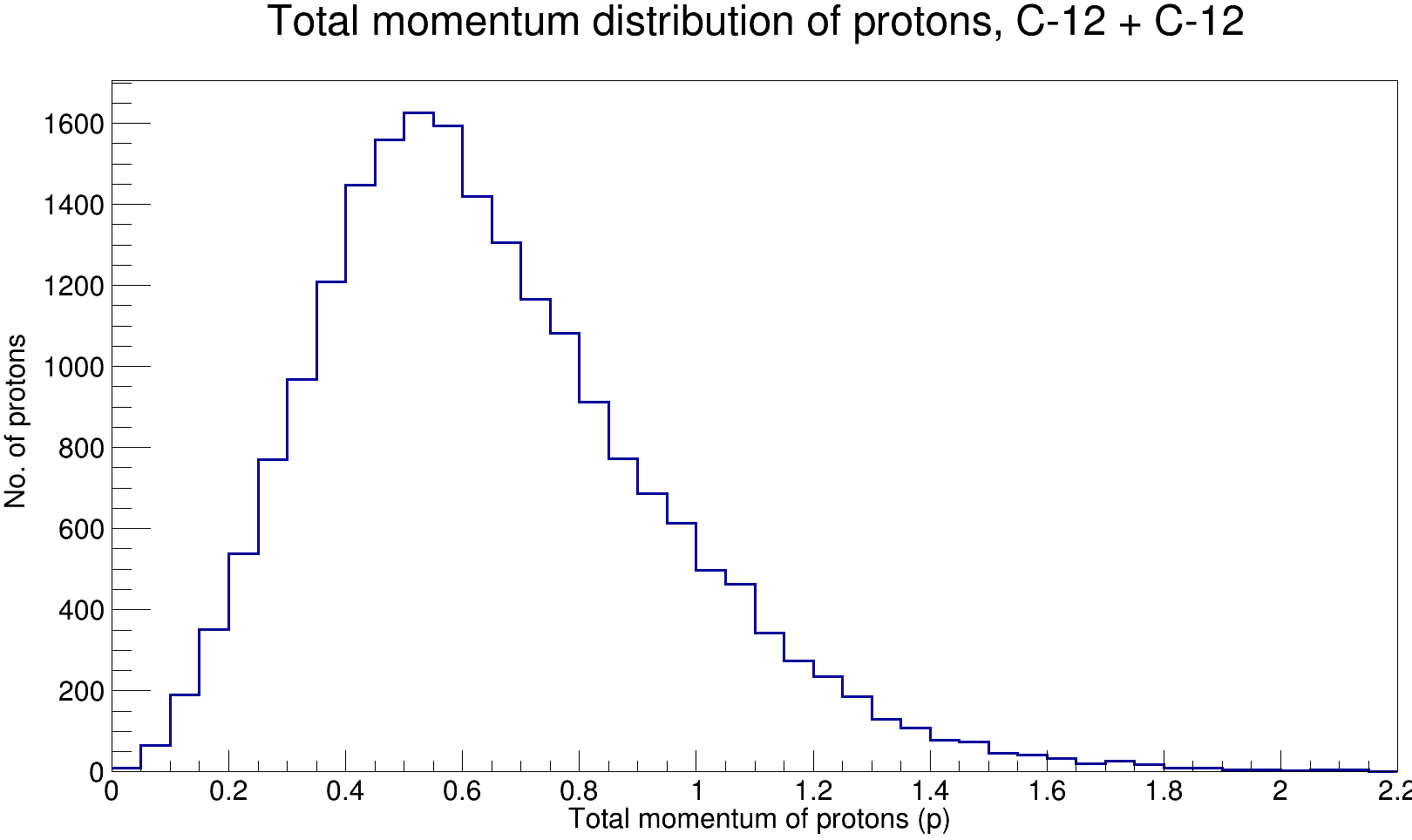}
\caption{p distribution of $p^{\pm}$ in $^{12}C-{^{12}C}$ collision.}
\label{Generator - Total momentum distribution of protons C12.}
\end{subfigure}
\hfill
\vspace{1cm}
\begin{subfigure}[h]{0.49\textwidth}
\centering
\includegraphics[scale=0.14]{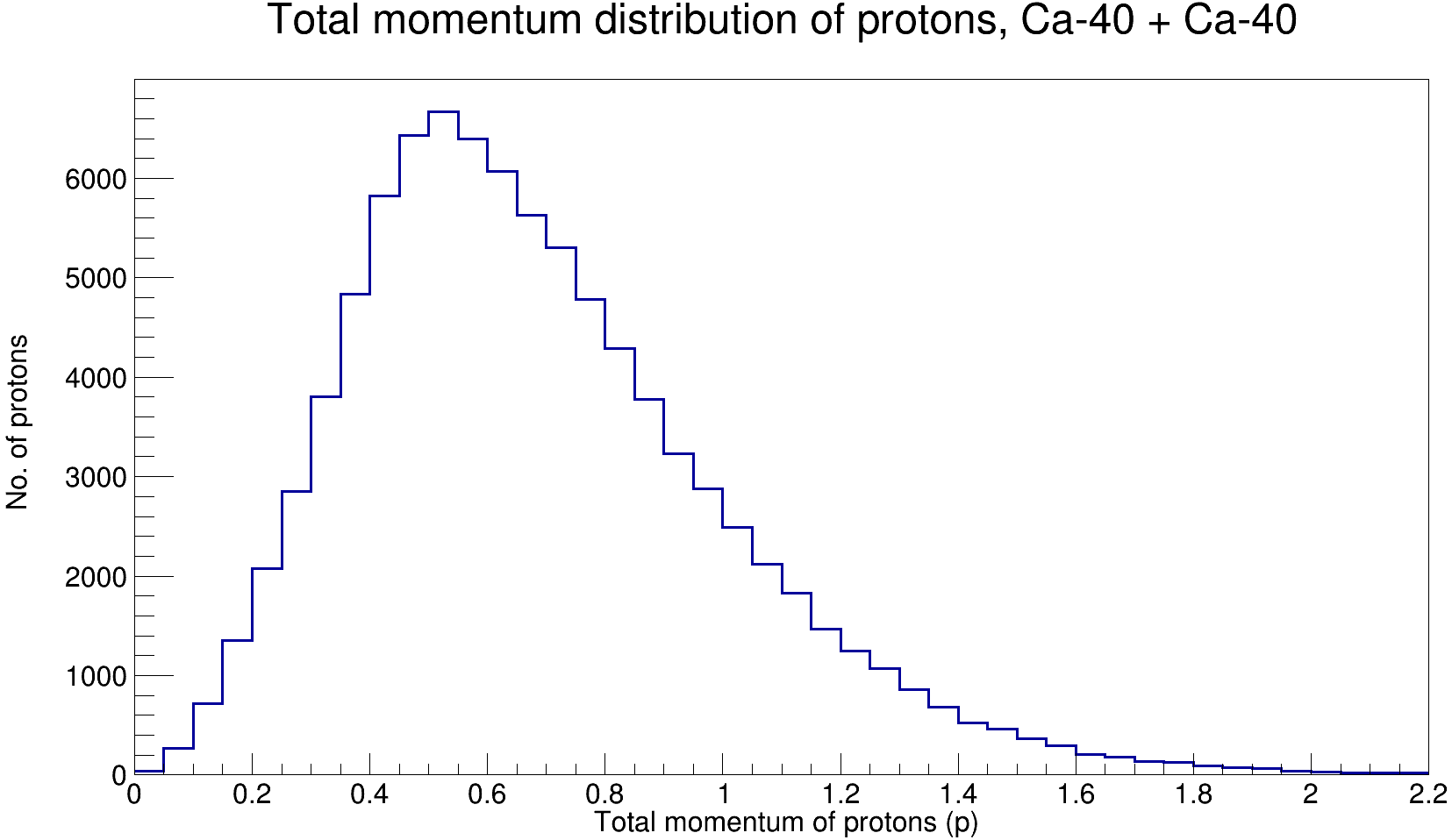}
\caption{p distribution of $p^{\pm}$ in $^{40}Ca-{^{40}Ca}$ collision.}
\label{Generator - Total momentum distribution of protons Ca40.}
\end{subfigure}
\hfill
\begin{subfigure}[h]{0.49\textwidth}
\centering
\includegraphics[scale=0.14]{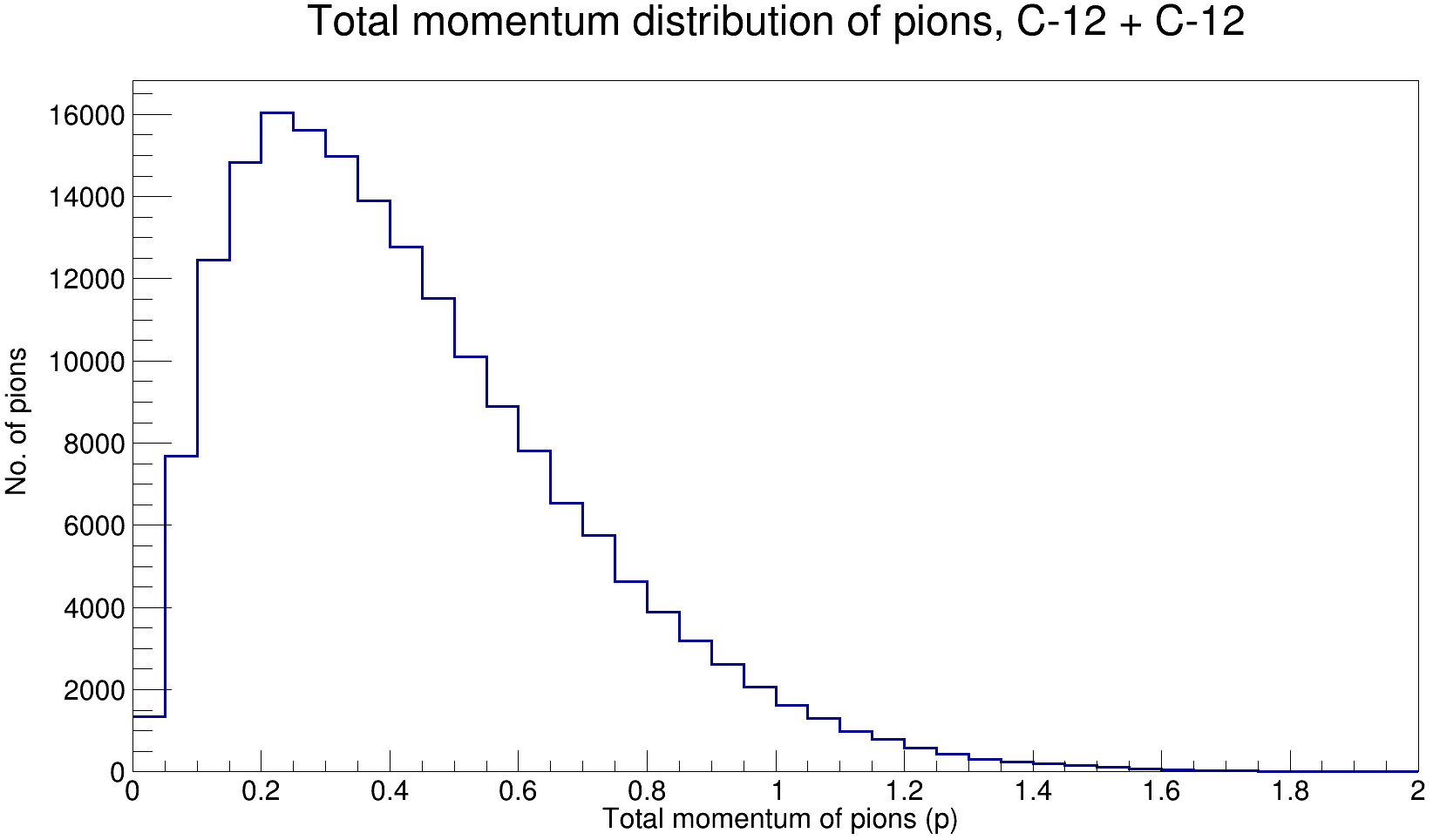}
\caption{p distribution of $\pi^{\pm}$ in $^{12}C-{^{12}C}$ collision.}
\label{Generator - Total momentum distribution of pions C12.}
\end{subfigure}
\hfill
\vspace{1cm}
\begin{subfigure}[h]{0.49\textwidth}
\centering
\includegraphics[scale=0.14]{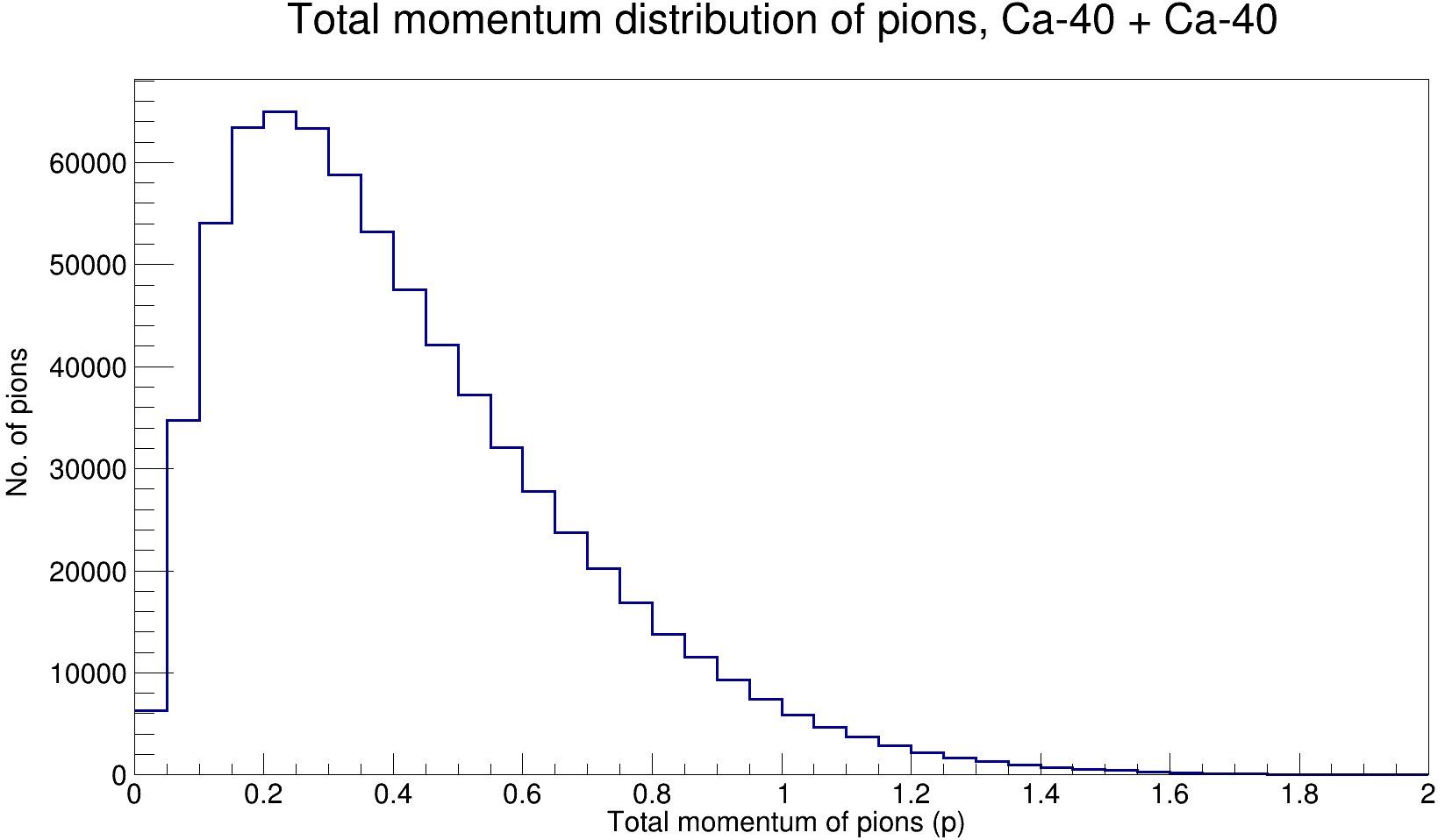}
\caption{p distribution of $\pi^{\pm}$ in $^{40}Ca-{^{40}Ca}$ collision.}
\label{Generator - Total momentum distribution of pions Ca40.}
\end{subfigure}
\hfill
\begin{subfigure}[h]{0.49\textwidth}
\centering
\includegraphics[scale=0.14]{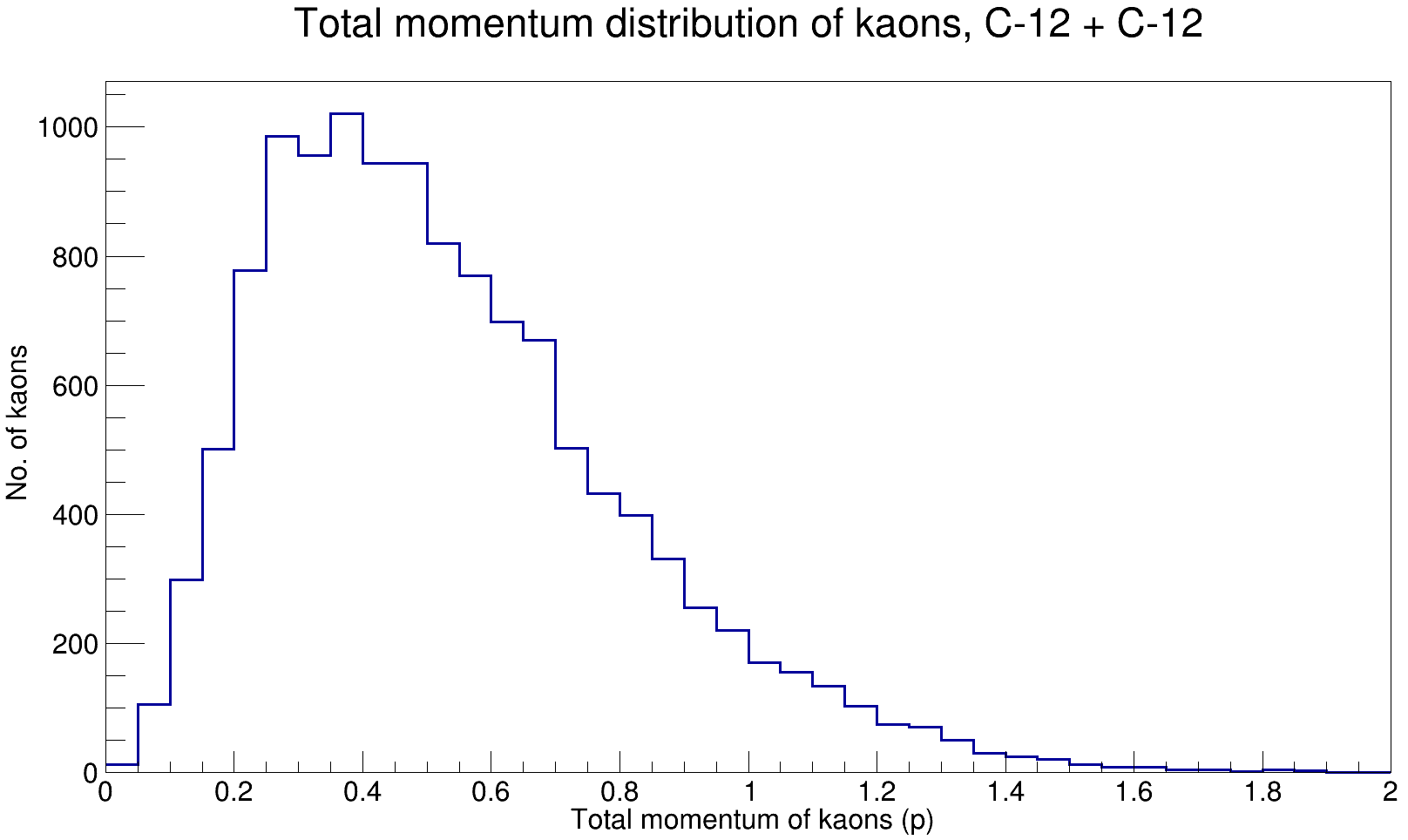}
\caption{p distribution of $k^{\pm}$ in $^{12}C-{^{12}C}$ collision.}
\label{Generator - Total momentum distribution of kaons C12.}
\end{subfigure}
\hfill
\begin{subfigure}[h]{0.49\textwidth}
\centering
\includegraphics[scale=0.14]{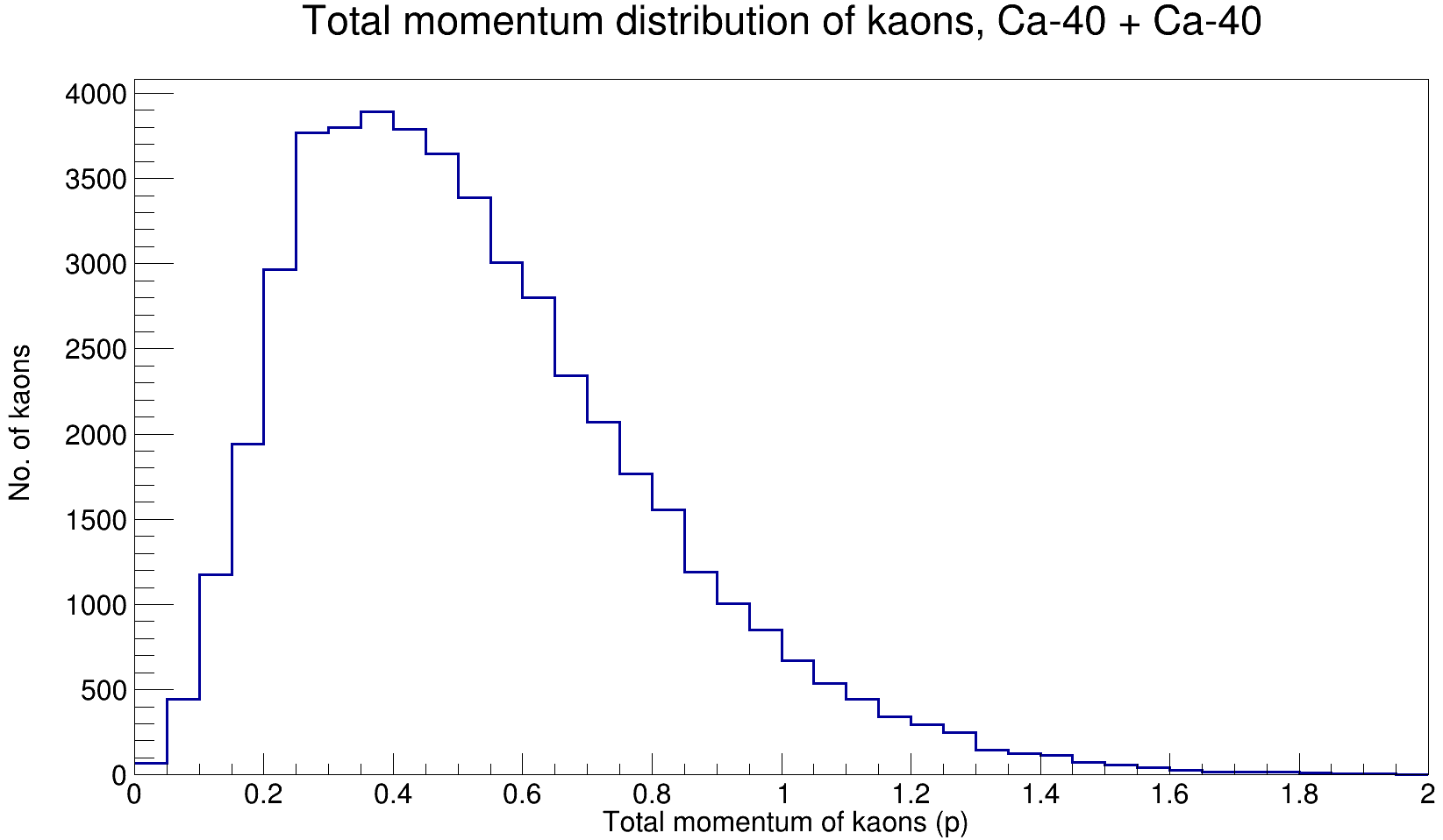}
\caption{p distribution of $k^{\pm}$ in $^{40}Ca-{^{40}Ca}$ collision.}
\label{Generator - Total momentum distribution of kaons Ca40.}
\end{subfigure}
\caption{Total momentum distribution of protons, pions, and kaons at generator level in $^{12}C-{^{12}C}$ and $^{40}Ca-{^{40}Ca}$ collision.}
\label{gen-p}
\end{figure}

Multiplicity of generated charged particles for $C-C$ and $Ca-Ca$ collisions are shown in
Fig.~\ref{gen-multipl}. The peaks at 12 for $^{12}C$+$^{12}C$ collisions and
at 40 for $^{40}Ca$+$^{40}Ca$ collisions correspond to events where no interaction occurred. The rapidity distributions are shown in Fig.~\ref{gen-rapidity}. The
spectra obtained from SMASH output show qualitative agreement with the ones
in Fig.~\ref{ABRAMOV_Paper_Fig.36}. Peaks for protons correspond to particles
moving close to the initial beam direction. Moreover, fractions of different particle
types can be estimated. It can be seen that for $|y|<2$ (i.e. within the acceptance
of the detector) charge particles are dominated by pions.  Apart from $p^{\pm},
\pi^{\pm},$ \& $K^{\pm}$, marginal numbers of sigmas, cascades, and omegas
were also generated. The PID efficiency depends on the particle momentum. The momentum
spectra for protons, pions, and kaons are shown in Fig.~\ref{gen-p} in the
midrapidity region ($|y|<0.5$ for which theoretical predictions has been given)  Most of
the pions have momentum below 0.8~GeV and protons~-~below 1~GeV. It means that
types of these particles should be well resolved by $dE/dx$ measurements.
When studying pion or proton spectra, there is high probability of kaon/pion
misidentification, but fraction of such events is strongly suppressed by small
initial kaon numbers.

\clearpage

\section{DETECTOR SIMULATION AND EVENT RECONSTRUCTION}
The detector simulation and reconstruction was performed with
the SpdRoot framework. To read SMASH generated events the
SpdRoot code was modified and additional C++ class was added.
During the simulation stage the particles
were transported through the detector geometrical model using
Geant4. At the reconstruction stage, Geant4 tracks and vertices
were reconstructed and particle identification with $dE/dx$ and
time of flight measurements was performed. For the PID three
hypotheses were considered: pion, kaon, and proton.
The reconstructed ionization energy losses and ``measured'' time of flight
were used to construct conditional probabilities (e.g. $p(t|pid)$,
where $t$ is the measured time and $pid$ is a particle type hypothesis). Out of 100K events generated by SMASH, first 1K events were considered for
detector simulation due to slow data processing in SpdRoot.

\section{ANALYSIS}
A physical analysis was performed using C++ codes and ROOT library
based on SpdRoot output. All tracks reconstructed in the detector with measured 
momentum were accepted. For the particle type the one that
gives the largest conditional probability is adopted. Multiplicity, as well as kinematic distributions for
pions, kaons, and protons are studied.
For particle momentum spectra there are no notable differences
between $C-C$ and for $Ca-Ca$ collisions, so only the first ones
will be considered.

\subsection{CHARGED TRACK MULTIPLICITY}
\label{CHARGED TRACK MULTIPLICITY, C12-C12}
\begin{figure}[h]
\centering
\begin{subfigure}[h]{0.49\textwidth}
\centering
\includegraphics[scale=0.14]{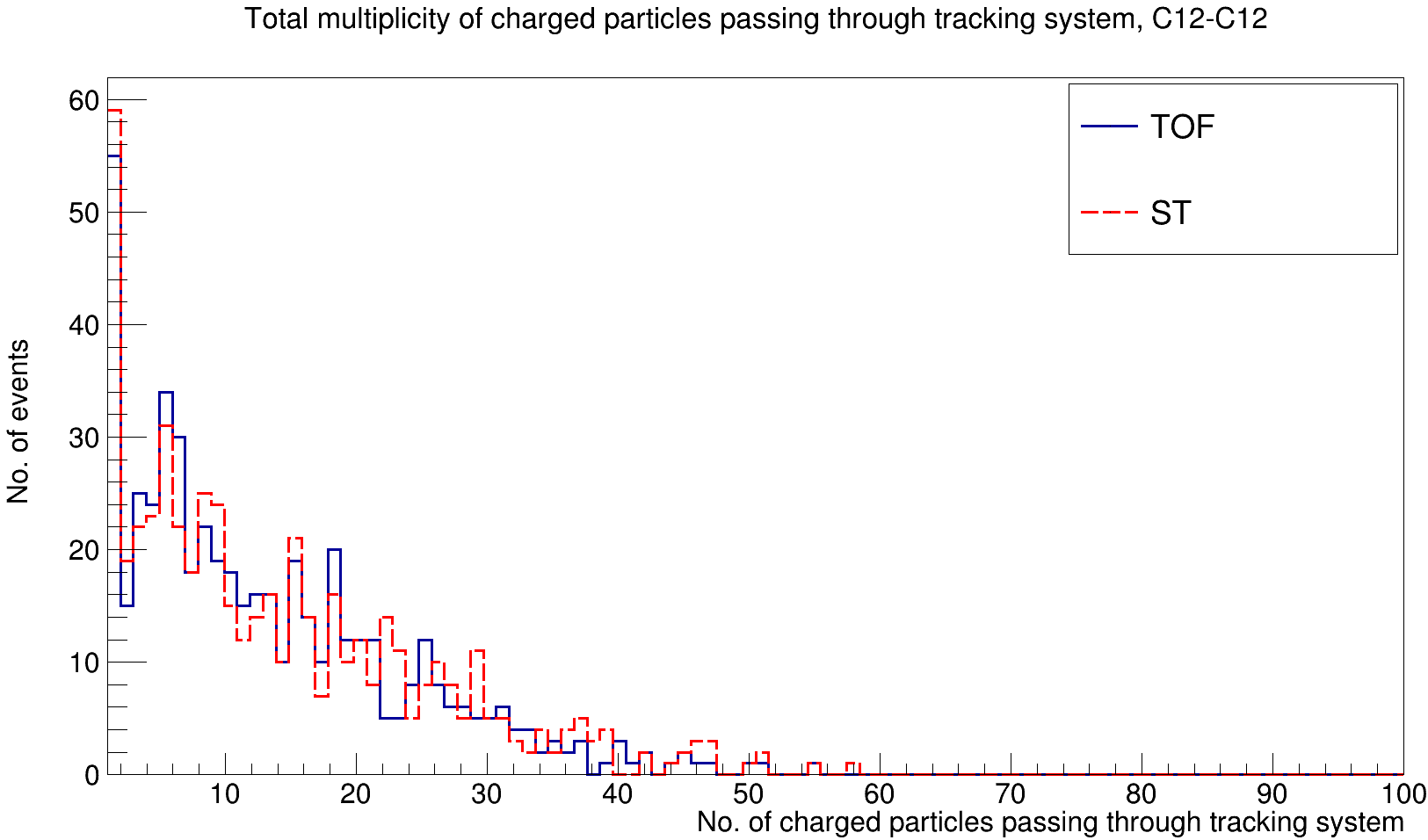}
\end{subfigure}
\hfill
\begin{subfigure}[h]{0.49\textwidth}
\centering
\includegraphics[scale=0.14]{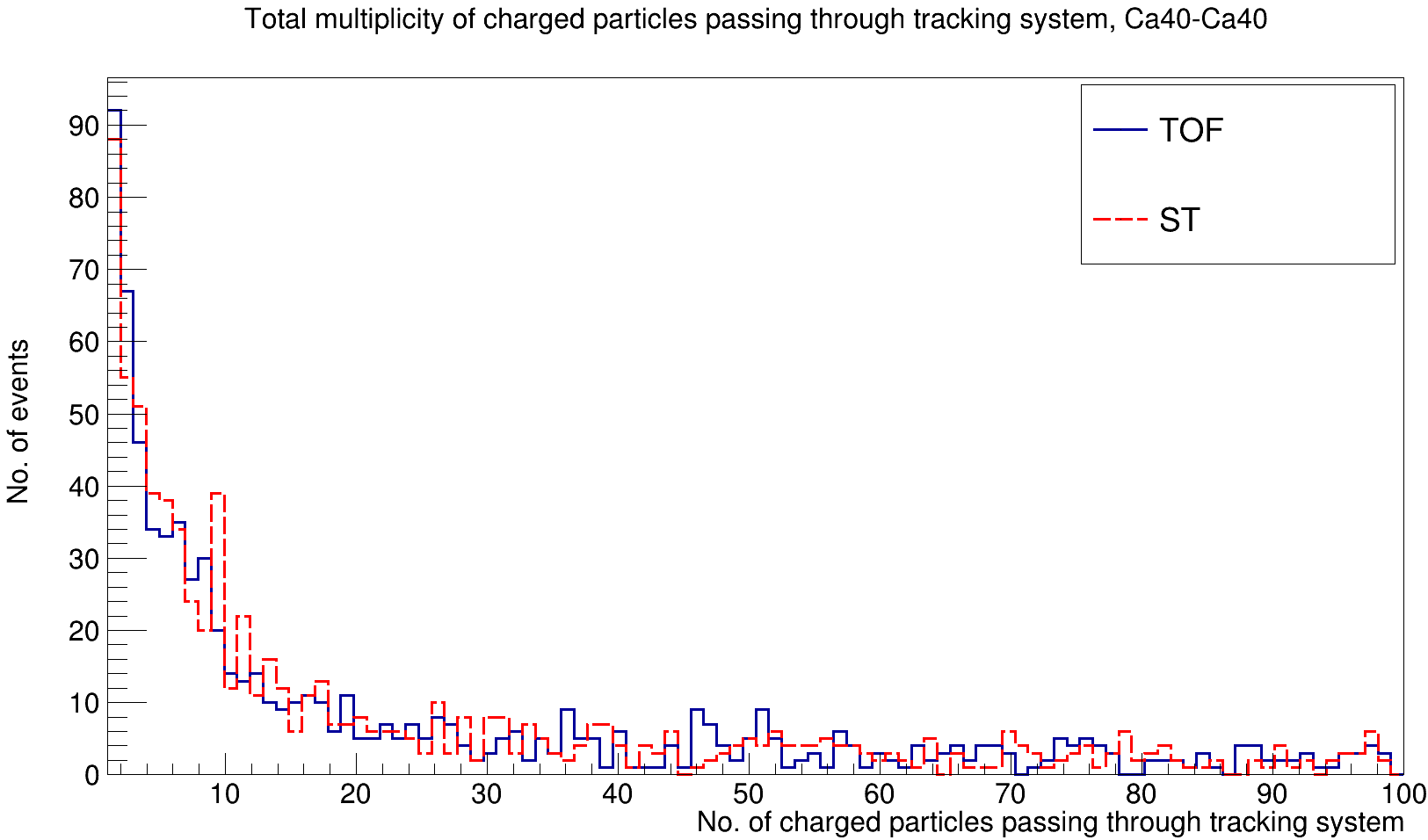}
\end{subfigure}
\caption{Charged track multiplicity reconstructed by in $^{12}C-{^{12}C}$ (left), $^{40}Ca-{^{40}Ca}$ (right) collisions (shown by red) and number of particles  for which TOF information
		is available (shown by blue).}
\label{sim-multiplicity-C-C and Ca-Ca}
\end{figure}

The SPD detector set-up is optimized for $p-p$ and $d-d$ collisions. Thus, knowing
charged track multiplicities for ion collisions is important to estimate
CT and ST occupancies and feasibility of such studies.
Fig.~\ref{sim-multiplicity-C-C and Ca-Ca} shows the total
multiplicity of charged particles reconstructed by the tracking system in
$^{12}C-{^{12}C}$ and $^{40}Ca-{^{40}Ca}$ collisions. 
The numbers of reconstructed tracks are much lower compared to generator-level
studies. It is because the geometry of the tracking system is such that, tracks with polar angle, $\theta < 10^{\circ}$ or $> 170^{\circ}$ do not hit the tracker and passes along the beam pipe itself, so such tracks are ignored. Also, there were events without nuclei interactions which resulted in no track reconstruction. So, to avoid a large peak at zero due to mentioned reasons, the X-axis count starts from 1. 


\subsection{PION MOMENTUM SPECTRUM ($^{12}C-{^{12}C}$)}
\begin{figure}[h]
\vspace{0.3cm}
	\centering
    \begin{subfigure}[h]{0.49\textwidth}
		\centering
		\includegraphics[scale=0.14]{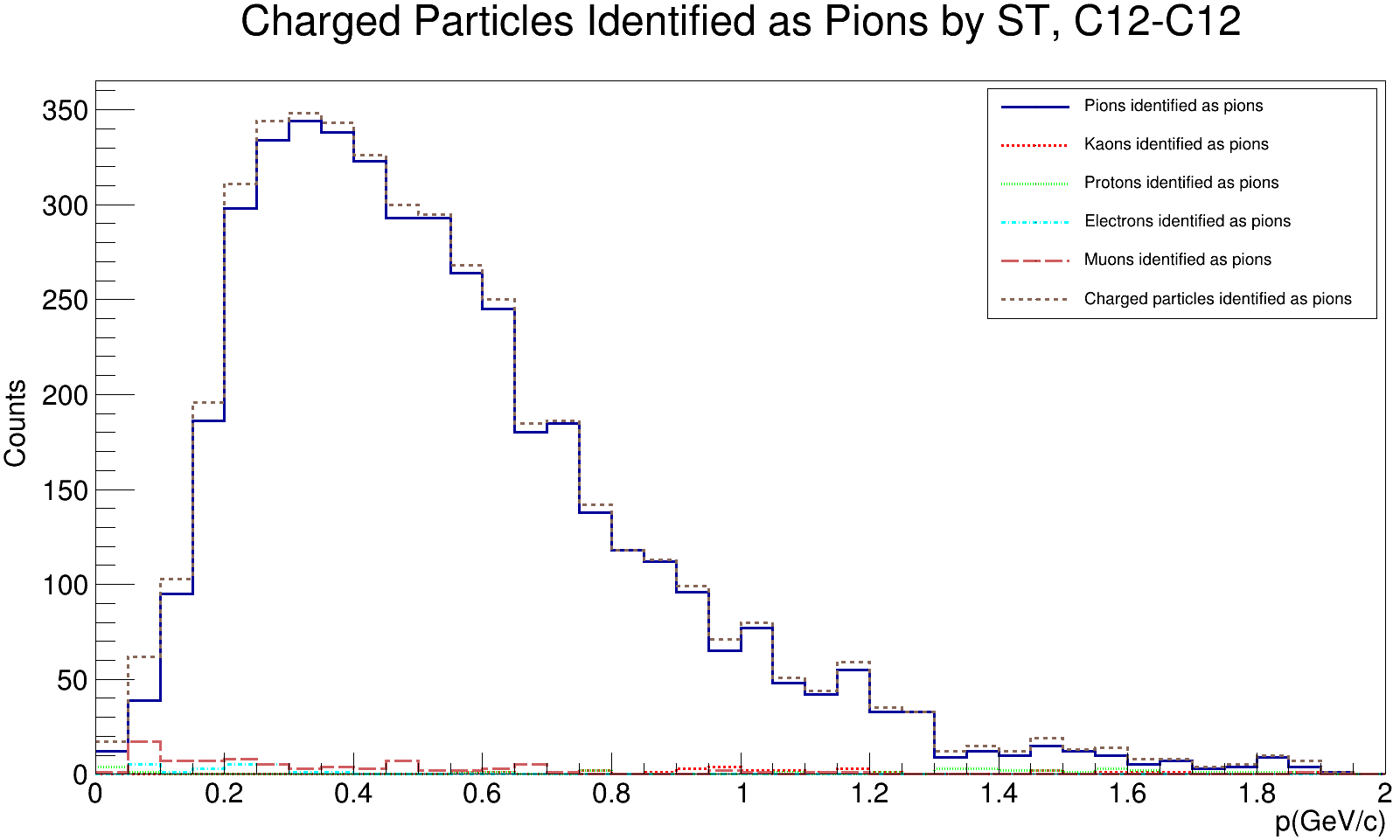}
		\caption{Total momentum distribution of reconstructed charged particles identified as $\pi^{\pm}$ by ionization losses.}
		\label{sim-p-pions-dedx}
	\end{subfigure}
	\hfill
	\begin{subfigure}[h]{0.49\textwidth}
		\centering
		\includegraphics[scale=0.14]{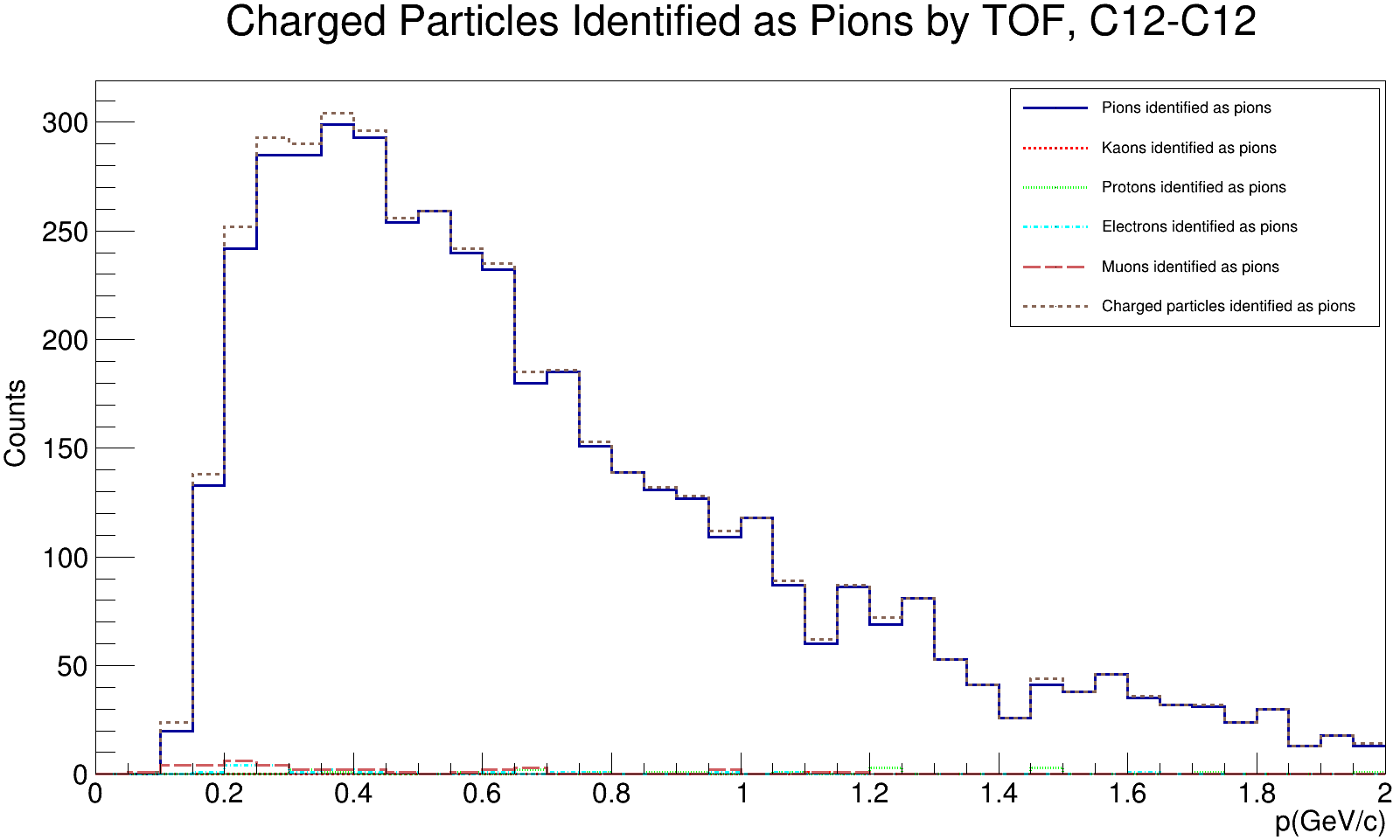}
		\caption{Total momentum distribution of reconstructed charged particles identified as $\pi^{\pm}$ by TOF.}
		\label{sim-p-pions-tof}
	\end{subfigure}
	\caption{Total momentum distribution of reconstructed $\pi^\pm$ candidates in $^{12}C-{^{12}C}$ collision (Detector level).}
	\label{sim-p-pions}
\end{figure}

\begin{figure}[h]
\vspace{0.3cm}
	\centering
    \includegraphics[width=.7\textwidth]{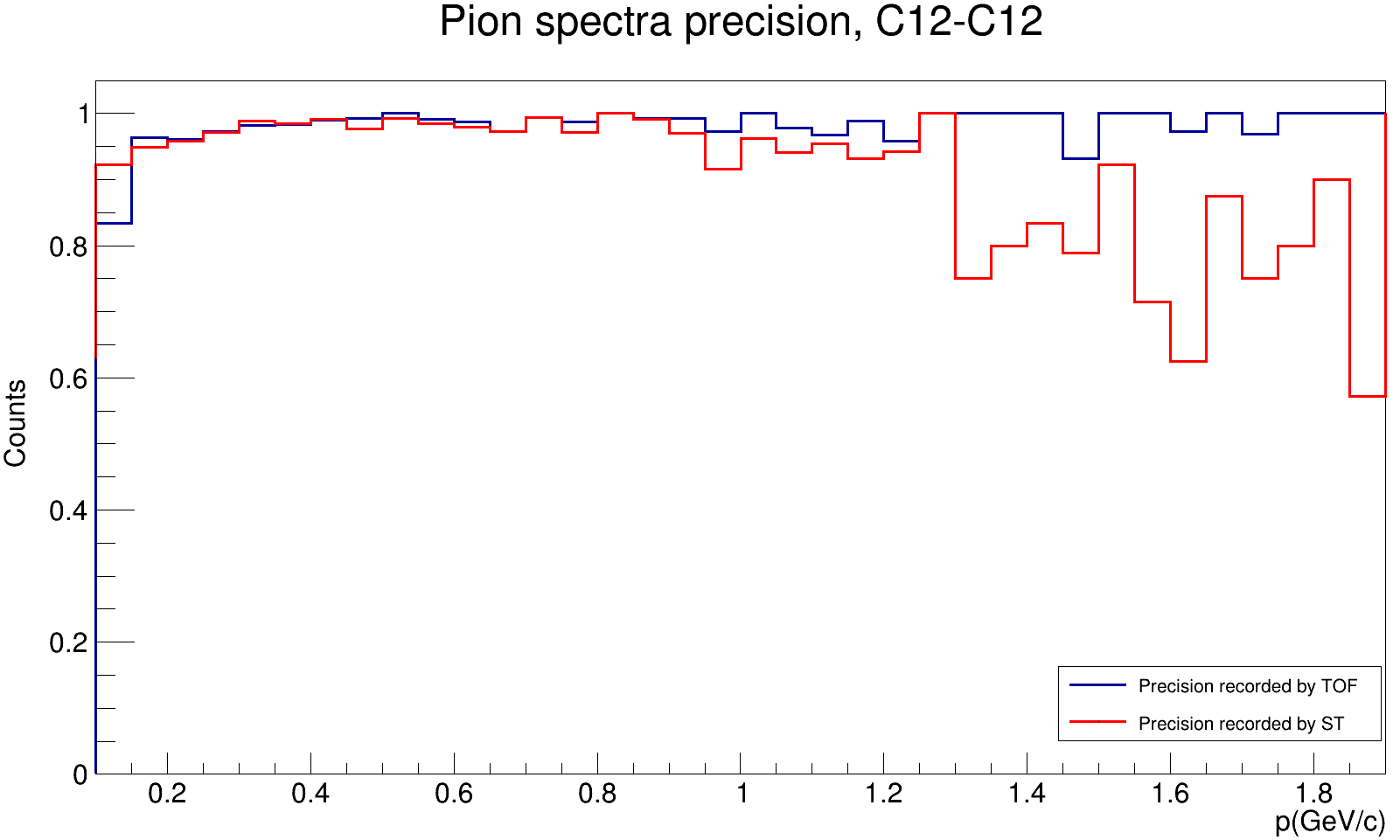}
	\caption{Purity of the selected pion candidates as a function of their momentum.}
	\label{pions-p-purity}
\end{figure}

The spectra of particles identified as pions separately by ionization losses
and by TOF are shown in Fig.~\ref{sim-p-pions} separately.
The spectra show resemblance with the generator plot of pion momentum distribution.
Based on MC-truth information backround from misidentification other charged particles
($K^{\pm}, p^{\pm}, e^{\pm},$ \& $\mu^{\pm}$) is studied. The obtained distribution
for ``pions identified as pions'' only slightly deviates from distribution of all
selected pion candidates. The estimated relative contamination of the pion spectra is
shown in Fig.~\ref{pions-p-purity}. It can seen that purity above 90\% can be obtained
up to 1.2~GeV using either $dE/dx$ or TOF measurements.

\clearpage

\subsection{KAON MOMENTUM SPECTRUM ($^{12}C-{^{12}C}$)}
\begin{figure}[h]
\vspace{1cm}
	\centering
	\begin{subfigure}[h]{0.49\textwidth}
		\centering
		\includegraphics[scale=0.14]{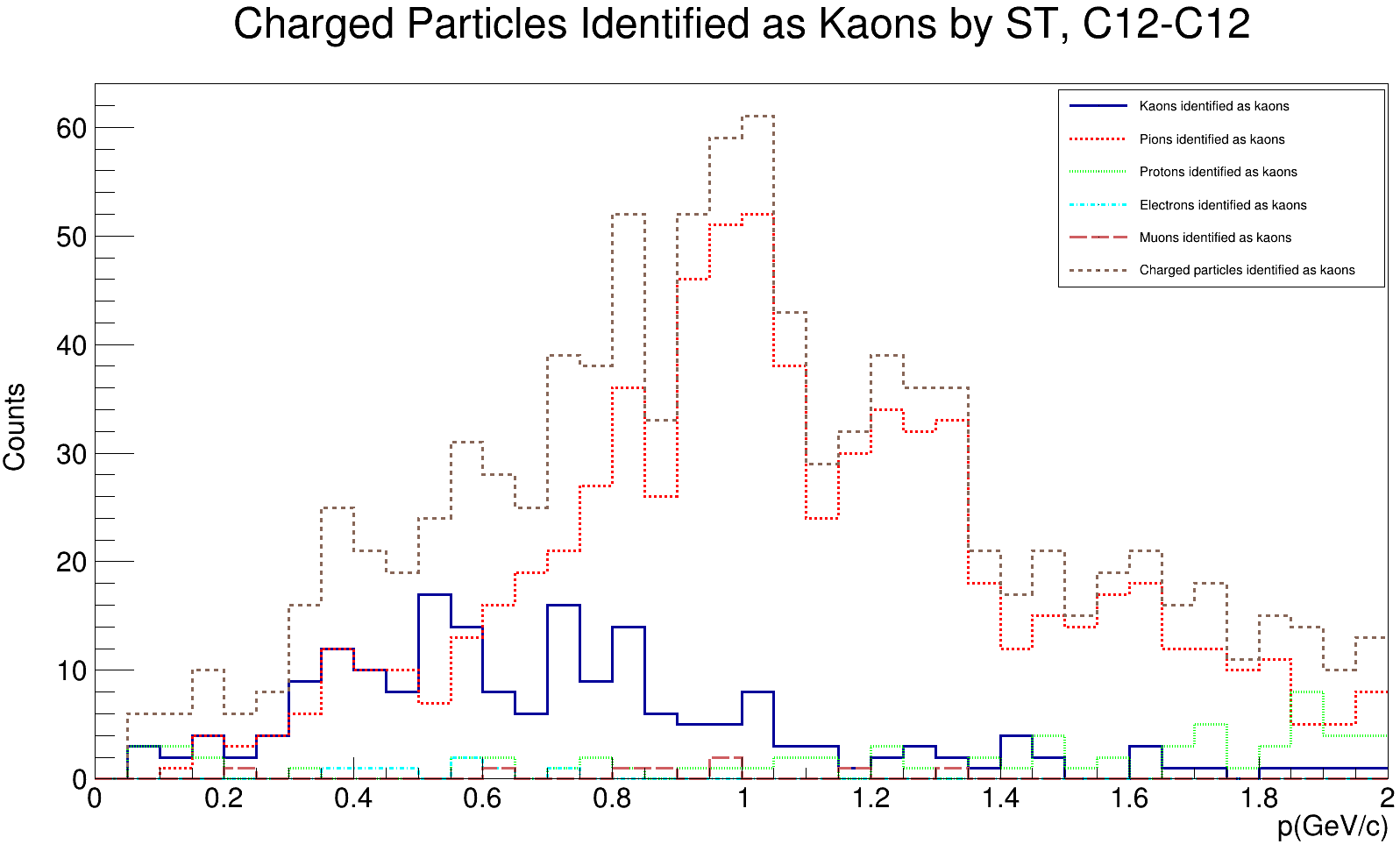}
		\caption{Total momentum distribution of reconstructed charged particles identified as $K^{\pm}$ by ionization losses.}
		\label{sim-p-kaons-dedx}
	\end{subfigure}
	\hfill
	\begin{subfigure}[h]{0.49\textwidth}
		\centering
		\includegraphics[scale=0.14]{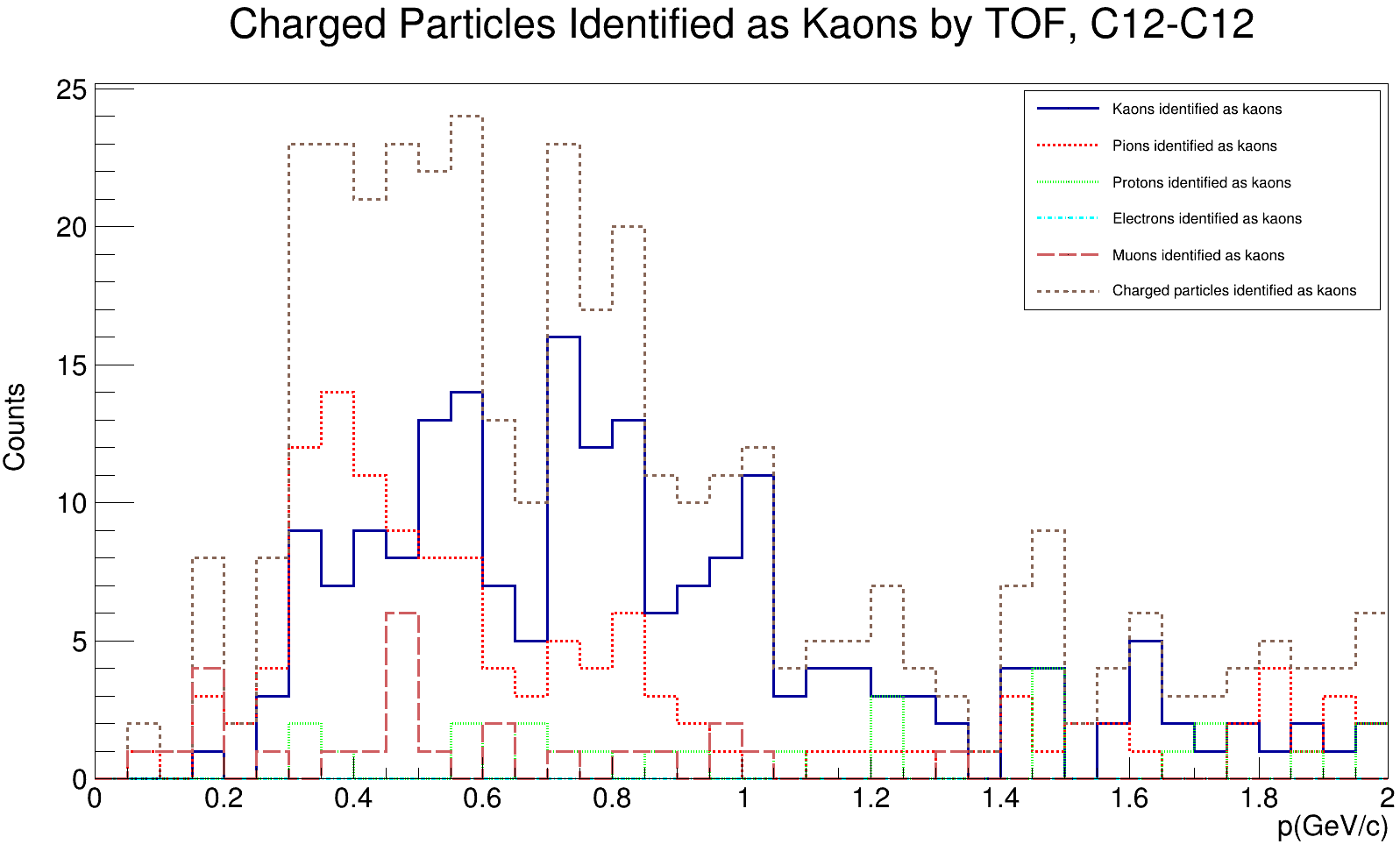}
		\caption{Total momentum distribution of reconstructed charged particles identified as $K^{\pm}$ by TOF.}
		\label{sim-p-kaons-tof}
	\end{subfigure}
	\caption{Total momentum distribution of reconstructed $K^\pm$ candidates in $^{12}C-{^{12}C}$ collision (Detector level).}
	\label{sim-p-kaons}
\end{figure}

\begin{figure}[h]
\vspace{0.1cm}
	\centering
	\includegraphics[width=.7\textwidth]{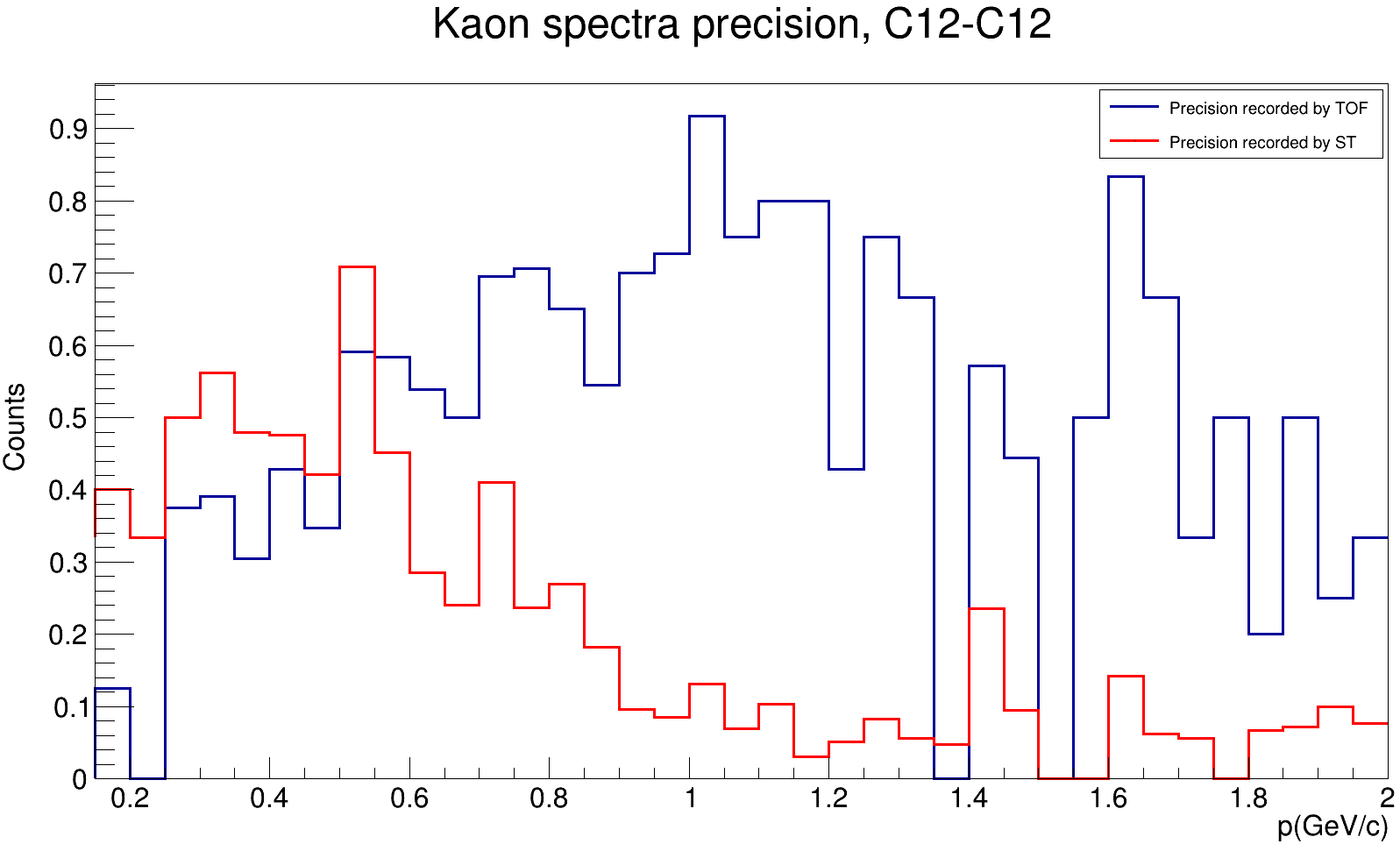}
	\caption{Purity of the selected kaon candidates as a function of their momentum.}
	\label{kaons-p-purity}
\end{figure}

The kaon momentum spectrum was explicitly mentioned among observables
to study hadron formation effects in nuclei. Nevertheless, kaon production may
be interesting for the reasons. The obtained spectra of kaon candidates is shown
in Fig.~\ref{sim-p-kaons} separately for ionization losses and TOF. First of all, the
shown data lack statistics. Secondly, it can bee seen that there is a huge contamination
from misidentified pions. This is explained by very small fraction of generated kaons
and the fact that probability to select misidentified particle is proportional to
their number. The relative fraction of correctly identified kaons in shown in Fig.~\ref{kaons-p-purity}.

\clearpage

\subsection{PROTON MOMENTUM SPECTRUM ($^{12}C-{^{12}C}$)}
\begin{figure}[h]
\vspace{1cm}
\centering
\begin{subfigure}[h]{0.49\textwidth}
	\centering
	\includegraphics[scale=0.14]{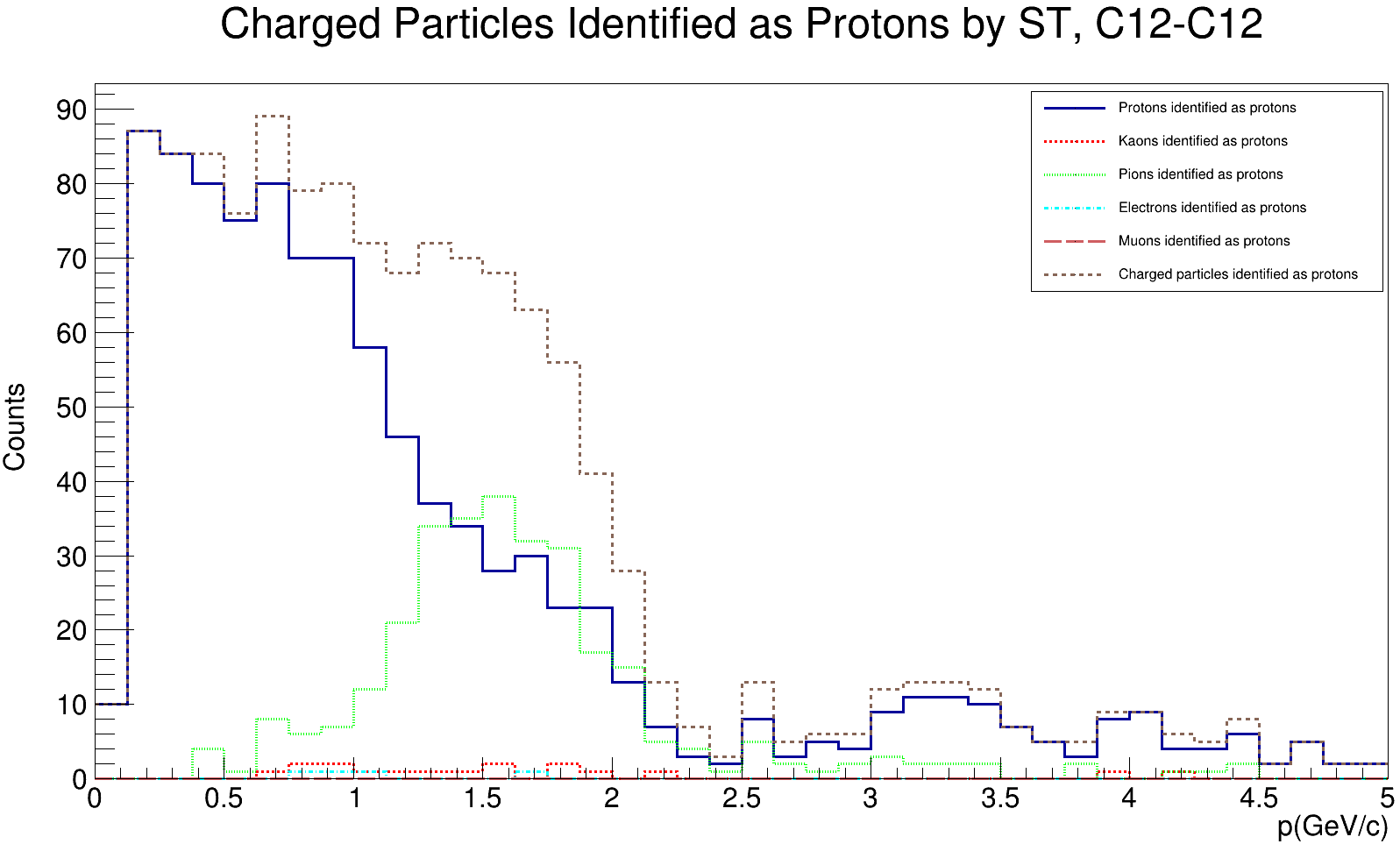}
	\caption{Total momentum distribution of reconstructed charged particles identified as $p^{\pm}$ by ionization losses.}
	\label{sim-p-protons-dedx}
\end{subfigure}
\hfill
\begin{subfigure}[h]{0.49\textwidth}
\centering
\includegraphics[scale=0.14]{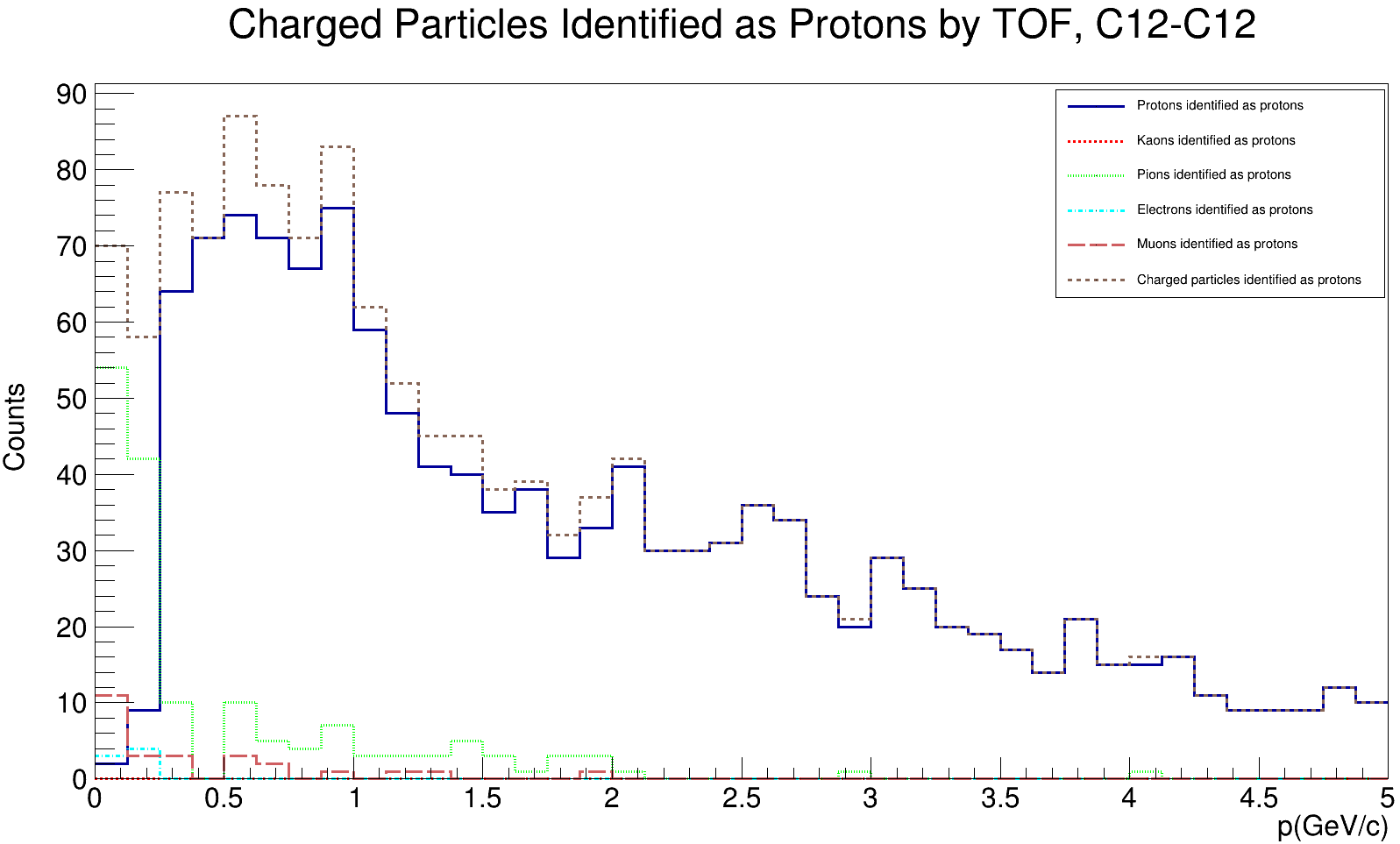}
\caption{Total momentum distribution of reconstructed charged particles identified as $p^{\pm}$ by TOF.}
\label{sim-p-protons-tof}
\end{subfigure}
\caption{Total momentum distribution of reconstructed $p^\pm$ candidates in $^{12}C-{^{12}C}$ collision (Detector level).}
\label{sim-p-protons}
\end{figure}

\begin{figure}[h]
\vspace{0.15cm}
	\centering
	\includegraphics[width=.7\textwidth]{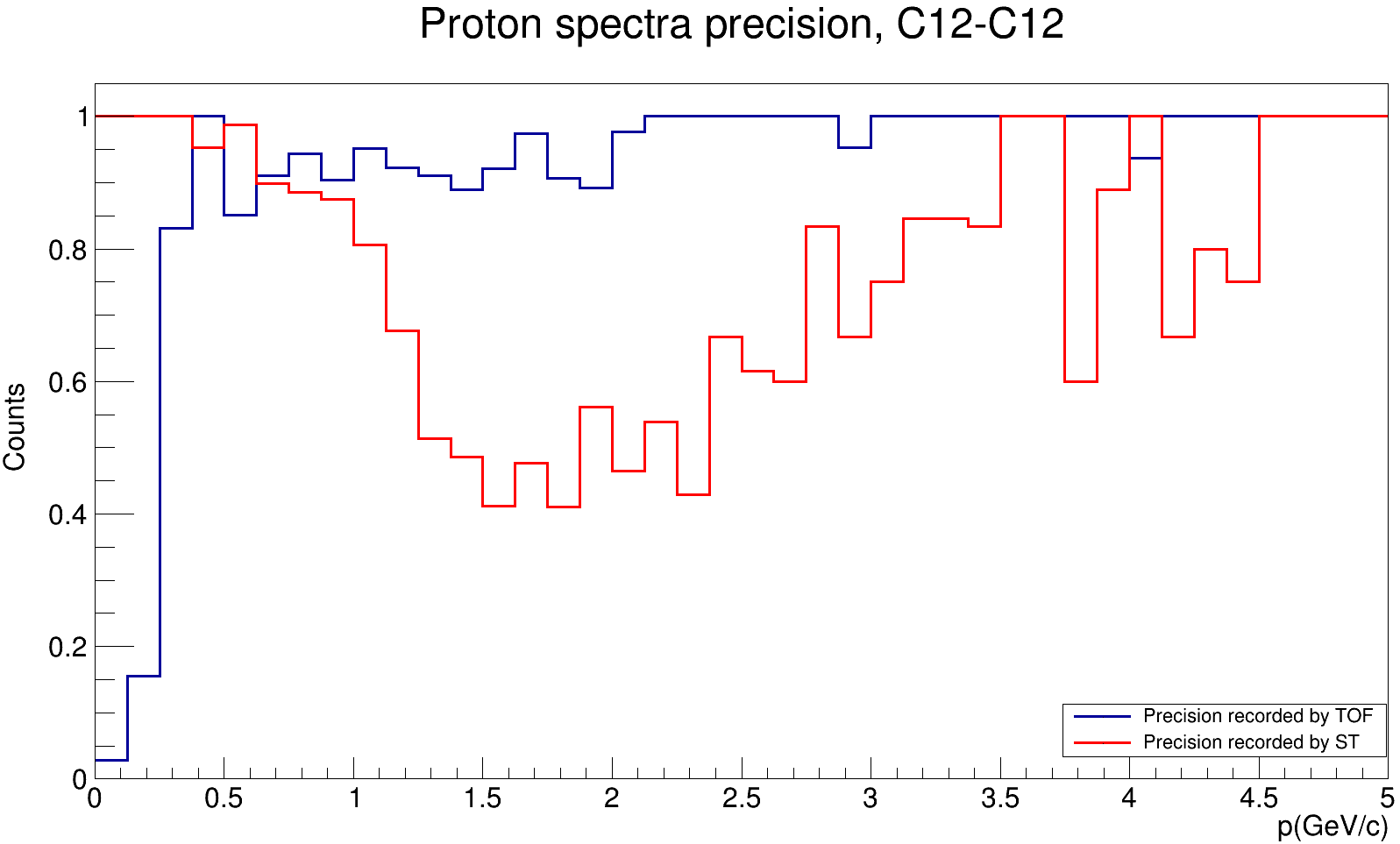}
	\caption{Purity of the selected proton candidates as a function of their momentum.}
	\label{protons-p-purity}
\end{figure}

Finally, proton momentum spectra have been considered. In this study
protons and antiprotons were considered together, but the fraction of produced
antiprotons is negligible. The proton candidate distributions and the contributions
from misidentification are shown in Fig.~\ref{sim-p-protons}. The purity of the
selected samples is shown in Fig.~\ref{protons-p-purity}. It can be seen
$dE/dx$ measurements alone will not allow precise determination of
proton spectrum. The reasonably good results can be expected only in case
of combined identification by ionization losses and TOF system.

\clearpage

\section{SUMMARY}
The goal of this work was to check the feasibility of hadron
formation effects studies at the first stage of SPD operation. For this purpose
an analysis of $^{12}C-{^{12}C}$ and $^{40}Ca-{^{40}Ca}$
collisions were performed at the generator level and then the full event reconstruction was done at detector level. The multiplicity distributions
indicate that occupancies of tracking detectors should be checks.
Part of the events with high number of charged tracks may not be
fully reconstructed. Particle identification with
ionization losses and TOF was considered separately (for
future $dE/dx$ only or their combination can be expected).
The purity of the measured charged pion distribution for both types of ion collisions using $dE/dx$ only is rather good and meets mentioned before requirements. In case of combination of information from ionization losses and time of flight system purity of proton distribution may be improved.




\begin{thebibliography}{9}
	\bibitem{Abramov:2021vtu}
	V.~V.~Abramov, A.~Aleshko, V.~A.~Baskov, E.~Boos, V.~Bunichev, O.~D.~Dalkarov, R.~El-Kholy, A.~Galoyan, A.~V.~Guskov and V.~T.~Kim, \textit{et al.}
	Phys. Part. Nucl. \textbf{52} (2021) no.6, 1044-1119
	doi:10.1134/S1063779621060022
	[arXiv:2102.08477 [hep-ph]].


\bibitem{SPDproto:2021hnm}
V.~M.~Abazov \textit{et al.} [SPD proto],
[arXiv:2102.00442 [hep-ex]].

\bibitem{TDR}
SPD TDR [unpublished].

\end{thebibliography}
\end{document}